\documentclass[12pt]{article}
\usepackage{amsmath,amssymb}
\usepackage{graphicx}
\usepackage[english]{babel}
\usepackage{cite}

\newcommand{\nc}{\newcommand}

\newcommand{\refb}[1]{(\ref{#1})}
\newcommand{\half}{\frac{1}{2}}

\newcommand{\be}{\begin{equation}}
\newcommand{\ee}{\end{equation}}
\newcommand{\bes}{\begin{equation*}}
\newcommand{\ees}{\end{equation*}}

\newcommand{\bea}{\begin{eqnarray}\displaystyle}
\newcommand{\eea}{\end{eqnarray}}

\nc{\NN}{{\cal N}}
\nc{\OO}{{\cal O}}
\nc{\KK}{{\cal K}}
\nc{\PP}{{\cal P}}
\nc{\UU}{{\cal U}}
\nc{\TO}{{\cal T}}
\nc{\KB}{{\cal K}}
\nc{\AN}{{\cal A}}
\nc{\MS}{{\cal M}}
\nc{\CC}{{\cal C}}
\nc{\DD}{{\cal D}}
\nc{\HH}{{\cal H}}
\nc{\GG}{{\cal G}}
\nc{\YY}{{\cal Y}}
\nc{\VV}{{\cal V}}
\nc{\II}{{\cal I}}

\nc{\BZ}{{\cal Z}}

\nc{\CY}{C{\rm{Y}_3}}
\nc{\ox}{\otimes}
\nc{\x}{\times}
\nc{\w}{\wedge}
\nc{\W}{\bigwedge}
\nc{\p}{\partial}
\nc{\bp}{\bar{\partial}}
\nc{\pbar}{\bar{\partial}}
\nc{\MM}{{\sf M}}
\nc{\wt}{\widetilde}
\nc{\wh}{\widehat}
\nc{\vp}{\varphi}
\nc{\ep}{\epsilon}
\nc{\vep}{\varepsilon}
\nc{\vtheta}{\vartheta}

\nc{\RR}{\mathbf{R}}
\nc{\RE}{{\cal R}}
\nc{\Z}{\mathbf{Z}}
\nc{\ZZ}{\mathbf{Z_2}}
\nc{\RP}{\mathbb{R}\mathbb{P}}
\nc{\cplx}{\mathbf{C}} % Complex Plane = Cplx Plane
\nc{\one}{{\mathbf{1}}}

\numberwithin{equation}{section}

\usepackage{babel}
\makeatother

\begin{document}

\rightline{HIP-2008-13/TH}

%\leftline\today\rightline\currenttime

\vskip 1cm \centerline{\large {\bf Disorder on the landscape}}
\vskip 1cm
\renewcommand{\thefootnote}{\fnsymbol{footnote}}
\centerline{{\bf Dmitry Podolsky,$^{1}$\footnote{dmitry.podolsky@helsinki.fi; \emph{URL}: www.nonequilibrium.net} Jaydeep Majumder,$^{3}$\footnote{majumder@mnnit.ac.in}%}} \centerline{{\bf
 and Niko Jokela$^{1,2}$\footnote{niko.jokela@helsinki.fi} }}
\vskip .5cm \centerline{\it ${}^{1}$Helsinki Institute of Physics
and ${}^{2}$Department of Physics } \centerline{\it
P.O.Box 64, FIN-00014 University of Helsinki, Finland} \centerline{\it ${}^{3}$Department of Physics,}
\centerline{\it Motilal Nehru National Institute of Technology,}
\centerline{\it Allahabad 211 004, India}

\setcounter{footnote}{0}
\renewcommand{\thefootnote}{\arabic{footnote}}

\begin{abstract}
Disorder on the string theory landscape may significantly affect dynamics of eternal inflation leading to the possibility for some vacua on the landscape to become dynamically preferable over others. We systematically study effects of a generic disorder on the landscape starting by identifying a sector with built-in disorder -- a set of de Sitter vacua corresponding to compactifications of the Type IIB string theory on %a class of
Calabi-Yau manifolds with a number of warped Klebanov-Strassler throats attached randomly to the bulk part of the Calabi-Yau. Further, we derive continuum limit of the vacuum dynamics equations on the landscape.
Using methods of dynamical renormalization group we %average over disorder and
determine the late %and late
time behavior of the probability distribution for an observer to measure a given value of the cosmological constant. We find the diffusion of the probability distribution to significantly slow down in sectors of the landscape where the number of nearest neighboring vacua for any given vacuum is small. We discuss relation of this slow-down with phenomenon of Anderson localization in disordered media.
\end{abstract}

\newpage

%%%%%%%%%%%%%%%%%%%%%%%%%%%%%%%%%%%%%%%%%%%%%%%%%%%%%%%%%%%%%%%%%%%%%%%%%%%%%%%%%%%
%%%%%%%%%%%%%%%%%%%%%%%%%%%%%%%%%%%%%%%%%%%%%%%%%%%%%%%%%%%%%%%%%%%%%%%%%%%%%%%%%%%
%%%%%%%%%%%%%%%%%%%%%%%%%%%%%%%%%%%%%%%%%%%%%%%%%%%%%%%%%%%%%%%%%%%%%%%%%%%%%%%%%%%
%%%%%%%%%%%%%%%%%%%%%%%%%%%%%%%%%%%%%%%%%%%%%%%%%%%%%%%%%%%%%%%%%%%%%%%%%%%%%%%%%%%

\tableofcontents

\section{Introduction and summary}

As we know, Universe is expanding with acceleration in the present
epoch of its evolution \cite{CCobservations}. It looks like that
1) acceleration becomes noticeable at redshifts $z\sim1$, and that 2) the
magnitude of acceleration itself does not strongly depend on the redshift
$z$ and is \emph{extremely} small in natural units. Indeed, the matter
density associated with this acceleration is of the order $10^{-29}\,{\rm g/cm}^{3}$,
$120$ orders of magnitude smaller than the Planckian density. It
remains unclear what is the meaning of this huge gap in the spectrum
of the underlying fundamental theory.

The physical nature of the source of this acceleration is also unclear.
In general relativity we have learned that the expansion of the universe filled with any known type of matter should
decelerate. Does Nature try to tell us that gravity itself should
be modified at $z\sim1$ \cite{ModifiedGravity} or is there simply
an additional degree of freedom \cite{ScalarField} with a very long
relaxation time and with a special equation of state $\epsilon_{\Lambda}\approx-p_{\Lambda}$?
This problem, arguably one of the most important one in the modern theoretical cosmology,
is known as the cosmological constant problem.

An attempt to resolve the cosmological constant problem stems from using
anthropic arguments \cite{Anthropic}. The logic behind anthropic reasoning is simple: universes (or Hubble patches)
%without supersymmetry and
with negative cosmological
constant, very rapidly collapse reaching an AdS (anti de Sitter) Big Crunch
singularity. Universes with large positive cosmological constant, on the other hand, expand so rapidly that the large scale structure does not
have enough time to form. In other words, observers able to measure
the value of cosmological constants do not appear in universes with
large positive and negative cosmological constants, and the bare fact
of our existence automatically implies that the value of the cosmological
constant should be small in our Universe. The statement that ``the cosmological constant is very small'' is true since if it were wrong there would be no observer to measure the cosmological constant.

We have to introduce a Multiverse (see, \emph{e.g.}, \cite{Weinberg:2005fh}) -- an immensely large collection
of Hubble patches with different values of cosmological constants. Following anthropic reasoning, the issue of the smallness of the cosmological constant of our Hubble patch is thus resolved.
%among them to ``explain'' the smallness of the cosmological constant
%in our Hubble patch, the one we live in.

An important argument in favor of the anthropic principle is that
such a system of Hubble patches can be realized on the string theory landscape of metastable vacua \cite{Susskind} populated by eternal inflation. %Still, even accepting rules of gameplay on the string theory landscape,
Even then, one might be left unhappy with an explanation of the cosmological constant's
smallness based on anthropic principle, since many questions remain unresolved. %within anthropic framework.
For example, it is still unclear how to properly define a gauge invariant probability for a given observer to measure a given value of the cosmological constant (recent discussion of the measure problem can be found in \cite{LindeMeasure,OtherMeasure}) or how to determine the distribution function of vacua within the string theory landscape.\footnote{The latter problem as well as the problem of calculating correlation functions weighted over such distribution function turns out to be NP-hard \cite{Douglas}.}

%In this paper, we do not adopt the anthropic principle. Instead, we try to
%establish a dynamical selection principle on the landscape that will automatically \emph{guarantee}
%that vacua with low positive values of the cosmological constant are
%preferable, independent whether observers exist or not in corresponding
%Hubble patches.

Instead of following the temptation of anthropic reasoning, one could try to establish a dynamical
vacuum selection principle on the landscape such that dynamics of the theory itself could automatically \emph{guarantee} that vacua with low positive values of the cosmological constant are
preferable.

One such dynamical selection principle on the landscape is based on
the analogue of the Anderson localization \cite{Mersini,Tye,PodolskyEnqvist}.
In condensed matter theory, Anderson localization was first discovered \cite{AndersonLocalization} as a phenomenon
of electron wave function localization inside a semiconductor, provided
that the disorder (for example, impurities or defects) of the effective
potential that electrons are experiencing is sufficiently strong.
Physics of Anderson localization in a semiconductor is related to existence of impurities,
called localization centers, with potential so strong that they bind
the Bloch wave function of the electron propagating in a disordered
medium.
%Mechanism responsible for the localization of the electron
%wave function near such impurities is interference between Bloch waves
%scattered by impurities or defects.

Since Anderson localization is a general phenomenon taking place in disordered media, it is plausible that with a significant amount of ``disorder'' on the string theory landscape may localize the wave function of the Universe in vacua with low cosmological constant \cite{Mersini,Tye}, thus providing an important dynamical selection principle.
In this paper, we qualitatively study the effects of disorder on the string theory landscape.

Instead of solving the Wheeler-De Witt equation with disordered potential \cite{Mersini,Tye} and in order to capture effects of eternal inflation important for populating vacua on the landscape we consider
%Garriga-Schwartz-Perlov-Vilenkin-Winitzki
vacuum dynamics equations \cite{VacDynEquations}
%
%
%I actually like my version more since it shows what is the difference between what we do and what Mersini and Tye
%do.
%
%\footnote{Another method is to analyze the Wheeler - De Witt equation with disordered potential %\cite{Mersini,Tye}.}
\be\label{eq:VacuumDynamicsNonVolume}
 \frac{dP_{i}}{d\tau}=\sum_{j}\left(\Gamma_{ji}P_{j}-\Gamma_{ij}P_{i}\right) = \sum_j H_{ij} P_j \ .
\ee
Equation (\ref{eq:VacuumDynamicsNonVolume}) describes eternal inflation and dynamics of tunneling between de Sitter vacua on the landscape. In this expression $\Gamma_{ij}$ are tunneling rates (inverse characteristic times of transition) between vacua $i$ and $j$. Due to %quantum mechanical
tunneling between different  vacua on the landscape, probabilities $P_{i}$ for an observer to measure a given value of the cosmological constant $\Lambda_{i}$ in her Hubble
patch (\emph{i.e.}, to find herself in a given vacuum $i$ on the landscape) evolve as time passes.
If the tunneling rates $\Gamma_{ij}$ have significant amounts of disorder, an analogue of Anderson localization may appear:
%If significant amount of disorder in the tunneling rates is
%present, an analogue of Anderson localization may appear, and
there may exist vacua, ``localization centers'',
such that the probability for an observer to live in such a vacuum
%find herself in corresponding
%Hubble patch
is higher. %than in other vacua.

The long time (large but finite $\tau$) behavior of the probability distribution function $P_i (\tau )$ is physically the most relevant to understand. We find it in three steps:
\begin{itemize}
 \item first we construct continuum limit of the vacuum dynamics equations (\ref{eq:VacuumDynamicsNonVolume}). This is possible when the number of vacua on the landscape is sufficiently large and tunneling rates $\Gamma_{ij}$ do not strongly fluctuate as functions of the position on the landscape (labeled by the index $i$).
 \item then we average over disorder. The reason for us to do this averaging is the following. When disorder is weak, one can treat its effects by means of perturbation theory (see Appendix \ref{sec:whyaverage}). Perturbed eigenstates and eigenvalues of the operator $H_{ij}$ in (\ref{eq:VacuumDynamicsNonVolume}) are given by the integrals over the moduli space of the landscape. Averaging over the moduli space of the landscape is equivalent to averaging over disorder, and the latter is technically much simpler to perform than the former.
 \item finally, we apply dynamical renormalization group (RG) methods to find the long (but finite) $\tau$ behavior of the distribution function $P_i(\tau)$ on a landscape with disorder.
\end{itemize}

The main result of our paper is that the diffusion of the probability
to measure a given value of the cosmological constant in a given Hubble
patch may indeed be suppressed in the parts of the landscape where
the number of nearest neighbors for any given vacuum is small.

This paper is organized as follows. We begin by identifying
a possible source of disorder on the string theory landscape in the Section \ref{sec:eternalKS}.
We argue that compactifying the Type IIB string theory on a special class of Calabi-Yau (CY) manifolds will induce disorder on the string theory landscape. We consider Calabi-Yau manifolds which have large numbers of both long and short Klebanov-Strassler (KS) throats.
%This source is due to the possibility of rather complex compactification of string theory on a Calabi-Yau space with multiple
%Klebanov-Strassler (KS) throats attached.
Disordering comes from the fact that the lengths of Klebanov-Strassler throats and
the points of their attachment to the bulk Calabi-Yau space may be essentially random.
As a byproduct, we show that it is possible to have eternal inflation of stochastic type in
the KKLMMT setup \cite{KKLMMT} with multiple Klebanov-Strassler throats.

In Section \ref{sec:continuum}
%and \ref{sec:Random-environment}
 we derive the continuum limit of the vacuum dynamics equations
(\ref{eq:VacuumDynamicsNonVolume}) by only using the fact that the number of vacua on the landscape (or a given sector of the landscape) is very large.
%and derive its continuum limit describing dynamics of eternal inflation
%on the landscape in terms of a diffusion equation with disorder. The latter can be considered as a Fokker-Planck %equation.
In Section \ref{sec:Disorder}
we study the continuum limit of (\ref{eq:VacuumDynamicsNonVolume}), which describes the dynamics of eternal inflation on a disordered landscape.
%in the particular case when only the anti-symmetric part of the tunneling rate is non-zero. This amounts to %studying diffusion in presence of ``random environment.''
In particular, we determine the late time (large but finite $\tau$) behavior of the probability distribution $P_{i}(\tau)$ by using dynamical renormalization group methods for various values of $N$ ---
effective dimensionality of
%the Hausdorff dimension of the tunneling graph corresponding to
the given island on the landscape. Section \ref{sec:conclusions} is devoted to discussion and conclusions.

%In section \ref{sec:Anisotropic-diffusion}, we study the dynamics of diffusion in presence of the symmetric part of %the tunneling rate. The presence of this term makes the diffusion random and anisotropic. We find that weak %diffusion in anisotropic diffusion coefficient is irrelevant for any value of $N$. In section %\ref{sec:HermitianCase}, we specialize to the isotropic case. In this case the Fokker-Planck equation can exactly %be solved in terms of the eigenfunctions of an associated Schrodinger equation with a supersymmetric effective %potential. For $N = 1$, the probability distribution shows the effect of Anderson localization at the local minima %of the effective potential.

%%%%%%%%%%%%%%%%%%%%%%%%%%%%%%%%%%%%%%%%%%%%%%%%%%%%%%%%%%%%%%%%%%%%%%%%%%%%%%%%%%%%%%%%
%%%%%%%%%%%%%%%%%%%%%%%%%%%%%%%%%%%%%%%%%%%%%%%%%%%%%%%%%%%%%%%%%%%%%%%%%%%%%%%%%%%%%%%%
%%%%%%%%%%%%%%%%%%%%%%%%%%%%%%%%%%%%%%%%%%%%%%%%%%%%%%%%%%%%%%%%%%%%%%%%%%%%%%%%%%%%%%%%
%%%%%%%%%%%%%%%%%%%%%%%%%%%%%%%%%%%%%%%%%%%%%%%%%%%%%%%%%%%%%%%%%%%%%%%%%%%%%%%%%%%%%%%%
%%%%%%%%%%%%%%%%%%%%%%%%%%%%%%%%%%%%%%%%%%%%%%%%%%%%%%%%%%%%%%%%%%%%%%%%%%%%%%%%%%%%%%%%

\section{Eternal inflation in a scenario with multiple \newline Klebanov-Strassler throats}\label{sec:eternalKS}

In this Section we apply the idea of vacuum dynamics to the inflaton field in string theory setup. We start with the setup as laid down in \cite{KKLMMT} where the open string modulus on a mobile D3-brane plays the role of the inflaton field. This class of models belong to the D3-anti-D3-brane inflation models. The D3/D7-brane model of inflation\cite{D3D7} is another model of inflation where one of the open string moduli plays the role of inflaton. There are also other models of inflation within string theory, where some other moduli, in particular closed string moduli, play the role of inflaton. Some specific models include K\"ahler moduli inflation\cite{Banks:1995dp,KahlerInflation}, complex structure moduli inflation\cite{CSMInflation}, racetrack models\cite{RacetrackInflation} and axion modulus inflation\cite{AxionInflation} and \cite{Misra} which used the $\alpha'$-correction to
K\"ahler potential\cite{Balasubramanian:2005zx} to produce a metastble
dS$_4$ vacuum instead of anti-D3-branes.

As mentioned before, we confine ourselves to the D3-anti-D3-brane inflation models. We consider a generic type IIB compactification on a Calabi-Yau orientifold in presence of flux. Locally at few places the geometry of the compactification is known: these local patches are denoted by the Klebanov-Strassler throats \cite{KS}. The D3-brane travel between various Klebanov-Strassler throats is equivalent to hopping %of the inflaton
between different de Sitter vacua within the string landscape. We are interested in showing how disorder gets introduced once the D3-brane starts traversing between various throats. This begs for an actual calculation of the tunneling rate of the inflaton field between various throats. The calculation of tunneling rate of the inflaton field from the first principle of string theory needs proper understanding of the dynamics of the inflaton field in the bulk Klebanov-Strassler geometry after incorporation of the back-reaction. We will discuss this aspect in a future publication \cite{PN}. Here we confine ourselves to the more qualitative feature of the problem.

We first show that it is possible for the inflaton field to reach the value within a single Klebanov-Strassler throat, where stochasticity takes over the classical motion. This is shown by proving the existence of a robust upper bound on the value of the inflaton field in such a throat. This helps us define a probability distribution of the inflaton field in over the stringy landscape of vacua. In the next step, we estimate the maximum possible number of larger Klebanov-Strassler throats for Calabi-Yau compactification in presence of flux.

\subsection{An upper bound on the inflaton field in a single-throat scenario}

In \cite{Baumann:2006cd}, it has been shown that the open string modulus which plays the role of inflaton, has a maximum range in warped D-brane inflation models. We will summarize their result here. In particular, we note that the upper bound on the open string modulus is independent of the details of the compact manifold and hence it is pretty much robust.

Let us assume a warped compactification to four spacetime dimensions in Type IIB string theory setup \cite{GKP}, with the ten-dimensional line element
\bea\label{10dansatz}
 ds^2_{10} & =      & e^{2A(r)}\eta_{\mu\nu}dx^{\mu}dx^{\nu} + e^{-2A(r)}(dr^2 + r^2 d\Omega_5) \\
           & \equiv & G_{MN}dx^Mdx^N \equiv G_{\mu\nu}dx^{\mu}dx^{\nu} + G_{mn}dy^m dy^n \\
           & \equiv & G_{\mu\nu}dx^{\mu}dx^{\nu} + e^{-2A}\widetilde{g}_{mn}dy^m dy^n \ ,
\eea
where $\eta_{\mu\nu}={\rm{diag}}(-1,1,1,1)$ and $\tilde g_{mn}$ is the metric on a six-dimensional Calabi-Yau manifold $\MS_6$. Our convention for the indices are $M,N=0,\ldots, 9$, $\mu,\nu=0,\ldots,3$ and $m,n=4,\ldots,9$. The radial coordinate $r$ of $\MS_6$ is defined by $r^2=(y^4)^2+\ldots +(y^9)^2$.

The internal space is a Klebanov-Strassler throat or deformed conifold which is a cone over a five-dimensional Sasaki-Einstein base manifold $\Omega_5$ \cite{KS}. This conical throat is assumed to be smoothly joined into the rest of the compact six-dimensional Calabi-Yau manifold $ \delta V_6^w$. The full metric over this compact Calabi-Yau is not known, however, locally at the throat it is known completely, {\em i.e.}, the warp factor $e^{2A}$ is known and it is a complicated function of $r$, closed string coupling $g_s$ and fluxes. The origin of the coordinates $r$ inside the throat is at the deformed tip of the conifold. We have turned on the following fluxes\footnote{We have also turned on the RR 0-form $C_0$.}
\be\label{eq:fluxes}
 \frac{1}{(2\pi)^2\alpha'}\int_A F_3 = M \quad ; \quad \frac{1}{(2\pi)^2\alpha'}\int_B H_3 = -K \ ,
\ee
where $M$, $K$ are integers, $A$ is the $S^3$ at the tip of KS throat, $B$ is a 3-cycle, Poincar\'e dual of $A$. In (\ref{eq:fluxes}) the field strengths $F_3=dC_2$ and $H_3=dB_2$, where $C_2$ and $B_2$ are the RR and NSNS two-forms, respectively; see, \emph{e.g.}, the reviews \cite{Douglas:2006es,Denef:2008wq} for more discussion.

\begin{figure}[!ht]
\begin{center}
\includegraphics[scale=0.5]{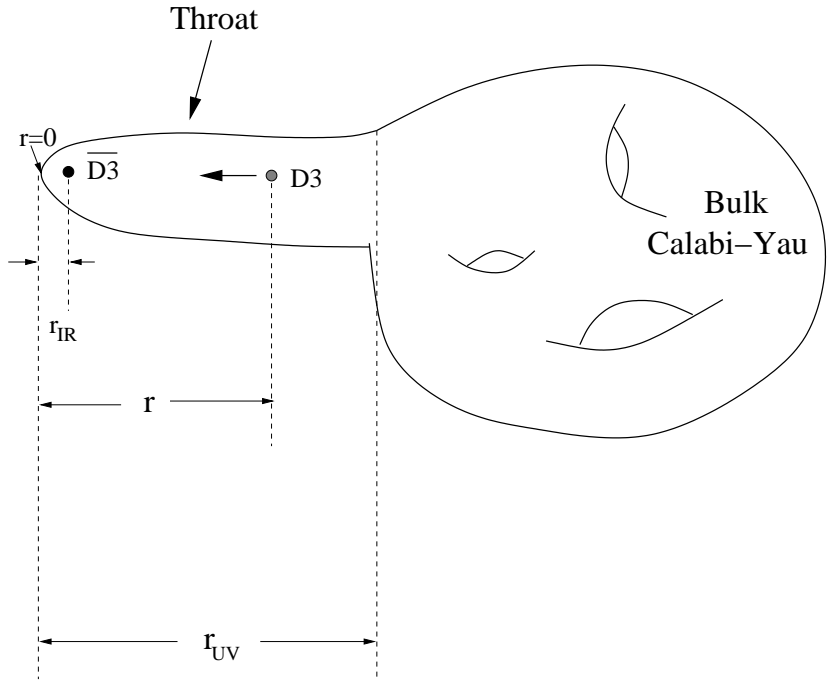}
\caption{The locations $r_{{\rm IR}}$ and $r$ of the D3 and anti-D3-brane, respectively, inside a single Klebanov-Strassler throat. It also shows the distance $r_{{\rm UV}}$ from the tip of the throat where the throat merges with the bulk Calabi-Yau. The origin $r = 0$ is located at the tip of the throat.\label{fig1}}
\end{center}
\end{figure}

Suppose, that the throat merges smoothly to the rest of the Calabi-Yau at some distance $r_{UV}$ measured from the tip. The anti-D3-brane is placed very near the tip, $r_{IR} \sim 0$, see Fig. \ref{fig1}. When $r_{IR} \ll r \ll r_{UV}$, the warped throat can be locally approximated as AdS$_5\times\Omega_5$, with the warp factor,
\be\label{adswarp}
e^{-4A} = \left(\frac{R}{r}\right)^4\,,
\ee
where $R$ is the radius of curvature of AdS$_5$. Since the background in \refb{10dansatz} is generated by fluxes of amount $MK$, we have the relation \cite{Gubser:1998vd}
\be\label{adscurvature}
 \frac{R^4}{\alpha'^2} = 4\pi g_s MK \frac{\pi^3}{{\rm{Vol}}(\Omega_5)}\ .
\ee
Here $\rm{Vol}(\Omega_5)$ denotes the dimensionless volume of the space $\Omega_5$ with unit radius.
Under the AdS approximation of the throat, we can think of the background as the near-horizon geometry of $N$ D3-branes, where $N = MK$.
%Note that the supergravity solution of the D3-brane is identical in both string and Einstein frames\cite{Horowitz:1991cd,Youm:1997hw}. This consideration will be helpful for calculations in future.

If $V^w_6$ denote the total warped volume of the compact six dimensional space, $\kappa_{10}$ the 10-dimensional
Newton constant, the four-dimensional Planck mass, $M_P$, is \cite{DeWolfe:2002nn}
\be\label{FourDimPlanck}
M_P^2 = \frac{V^w_6}{g_s^2\kappa^2_{10}} \ ,
\ee
where $\kappa^2_{10} = \half(2\pi)^7 (\alpha')^4$. The warped volume of the internal space is
\be\label{warpedvolume}
V^w_6 = \int d^6 y \sqrt{\tilde{g}_6}\ e^{-4A} \ ,
\ee
where $\widetilde{g}_6 = \det \widetilde{g}_{mn}$.

We can formally split the volume of the internal space into the volume of the throat, $V^w_{{\rm throat}}$,  plus the rest of the Calabi-Yau, $\delta V_6^w$,
\be\label{CYVolume}
V^w_6 = V^w_{{\rm throat}} + \delta V^w_6 \ .
\ee
Under AdS approximation, the throat contribution is
\bea\label{throatvolume}
V^w_{{\rm throat}} & = & {\rm{Vol}}(\Omega_5)\int_0^{r_{UV}} dr r^5 e^{-4A(r)} \nonumber\\
                   & = & \half {\rm{Vol}}(\Omega_5) R^4r^2_{UV} \nonumber\\
                   & = & 2\pi^4 g_s MK\alpha'^2 r^2_{UV} \ ,
\eea
where in the last step, we have used the relation \refb{adscurvature}. Note that \refb{throatvolume} is independent of ${\rm Vol}(\Omega_5$). Since the bulk volume is model-dependent and we are looking for a fairly robust upper bound of the inflaton field, we use the fact $V_6^w > V^w_{{\rm throat}}$, so that we can write \refb{FourDimPlanck} as
\be \label{MP}
M_P^2 > \frac{V^w_{{\rm throat}}}{g_s^2\kappa^2_{10}} \ .
\ee
If $r$ denotes the mobile D3-brane radial modulus with respect to the anti-D3-brane inside the throat, the canonically normalized inflaton field is $\phi = \sqrt{T_3} r$, where $T_3 = 1/\left((2\pi)^3\alpha'^2 g_s\right)$ is the physical D3-brane tension. The maximum allowed value of the inflaton field is thus $\phi_{{\rm max}} = \sqrt{T_3}r_{{\rm UV}}$. Hence we can write
\be\label{eq:premaximum}
\left(\frac{\phi_{{\rm max}}}{M_P}\right)^2 \simeq \frac{T_3 r^2_{{\rm UV}}}{M_P^2}
< \frac{T_3g_s^2\kappa^2_{10}r^2_{{\rm UV}}}{V^w_{{\rm throat}}} \ .
\ee
Using \refb{adscurvature} and the values of $T_3$ and $\kappa_{10}$ in (\ref{eq:premaximum}), we get the maximum allowed value of the inflaton field in four-dimensional Planck units, in a
Klebanov-Strassler throat \cite{Baumann:2006cd}
\be\label{maximum}
 \left(\frac{\phi_{{\rm max}}}{M_P}\right)^2  < \frac{4}{MK} \ .
\ee
Thus, for example, for $MK = 100$, $\phi_{{\rm max}} \simeq 0.2 M_P$.

Now we should find whether it is possible to have eternal inflation with
$\phi_{{\rm max}} \simeq 0.2 M_P$ within this setup. The distance $r$ between the D3 brane and the anti-D3 brane
after an appropriate canonical normalization $\phi=\sqrt{T_{3}}r$
plays the role of the inflaton in the four-dimensional effective field
theory. The effective potential for the inflaton has the form
\begin{equation}
 V(\phi)=\frac{4\pi^{2}\phi_{0}^{4}}{MK}\left(1-\frac{\phi_{0}^{4}}{MK\phi^{4}}\right) \ , \label{eq:KKLMMTpotential}
\end{equation}
where $\phi_0 = \sqrt{T_3}r_0$ denotes the location of anti-D3-brane. Around $\phi_0$, \emph{i.e.}, near the tip of the throat, the geometry deviates from AdS$_5\times S^5$. From \refb{eq:KKLMMTpotential}, we get the slow-roll parameters
\bea
\epsilon(\phi) & = & \frac{M_{P}^{2}}{2}\left(\frac{V'}{V}\right)^{2}\approx\frac{8}{(MK)^{2}}
\frac{M_{P}^{2}\phi_{0}^{8}}{\phi^{10}}  \label{eq:KKLTepsilon} \\
 \eta(\phi) & = & M_{P}^{2}\frac{V''}{V} \approx -\frac{20}{MK}\frac{M_{P}^{2}\phi_{0}^{4}}{\phi^{6}} \ .\label{eq:KKLTeta}
\eea
If the conditions $\epsilon,\eta\ll1$ are satisfied, then we have
an inflationary dynamics in our setup. Among the two conditions, \refb{eq:KKLTeta}
is the more restrictive one. From there, we have
\begin{equation}
\phi\gg\left(\frac{20}{MK}M_{P}^{2}\phi_{0}^{4}\right)^{1/6} \ . \label{eq:KKLTslowrollphi}
\end{equation}
Therefore, if the vacuum expectation value of $\phi$ is large enough, then we have inflationary dynamics.

During one Hubble time $\Delta t\sim H^{-1}$ the value of the inflaton
field decreases by
\begin{equation}
\Delta\phi\sim\dot{\phi}\Delta t\sim\frac{1}{3H^{2}}\frac{\partial V}{\partial\phi}\sim\frac{M_{P}^{2}}{8\pi V}\frac{\partial V}{\partial\phi}
\sim \frac{M_P}{8\pi}\sqrt{2\epsilon(\phi)}= \frac{M_{P}^{2}}{2\pi MK}\frac{\phi_{0}^{4}}{\phi^{5}} \ .
\label{eq:Deltaphiclass}
\end{equation}
Now let us take the quantum fluctuations of the inflaton field into account.
During the same Hubble time the quantum fluctuations $\delta\phi$ of the
field $\phi$ are generated; for the fluctuations with the wave length
$l\sim k^{-1}\sim H^{-1}$ we have the following estimation for the
amplitude
\begin{equation}
|\delta\phi|\sim\frac{H}{2\pi}\sim\sqrt{\frac{2V}{3\pi M_{P}^{2}}}
\sim\sqrt{\frac{8\pi\phi_{0}^{4}}{3M_{P}^{2}MK}}=\sqrt{\frac{8\pi}{3MK}}\frac{\phi_{0}^{2}}{M_{P}} \ . \label{eq:Deltaphiquant}
\end{equation}
Therefore, if $\phi$ reaches the value
\begin{equation}\label{eternal}
 \phi_{stoch}\sim\left(\frac{M_{P}^{3}\phi_{0}^{2}}{10\sqrt{MK}}\right)^{1/5}\sim10^{-2}M_{P}
\end{equation}
or higher, the stochastic force becomes more important than the classical
and the inflation enters the eternal regime \cite{StochasticInflation}. Comparing \refb{maximum} with \refb{eternal}, we find that the maximum value of the inflaton field allowed by string theory in a single Klebanov-Strassler throat, is at least ten times bigger in magnitude than what is needed to trigger the stochastic force.\footnote{According to \cite{Chen:2006hs}, the maximum possible value $\phi_{\rm max}$ of the inflaton field in brane inflation scenarios with warping is such that regime of eternal inflation is not allowed. The estimation
of $\phi_{\rm max}$ by X. Chen et al. is based on the rough estimation of the volume of the Calabi-Yau as $V_6^w \sim L^6$ and subsequent requirement that $R < L$ (as follows from the value of the $4$-dimensional Planckian mass). The estimation of the $CY_6$ volume (\ref{throatvolume}), (\ref{MP}) taking warping of the throat into account is more accurate.}
Since the bound obtained in \refb{maximum}
is much robust and independent of the details of the throat, we conclude that it is possible to have eternal inflation even with a single Klebanov-Strassler throat.

Finally, let us note that the regime of eternal brane inflation is in a sense opposite to the one usually considered in literature (see for example \cite{BraneReheating} and references therein): regime of deterministic slow-roll or DBI inflation corresponding to relatively small values of $\phi$ and ending with a D3-anti-D3 pair annihilation at the tip of the throat. Sufficiently short KS throats where D3-brane quickly approaches the tip of the throat effectively play a role of sinks on the multithroat sector of the landscape that we consider. Once the D3-brane enters such a throat, eternal inflation comes to the end for an observer living on the brane.

\subsection{An upper bound on the number of large \newline Klebanov-Strassler throats}\label{sec:numthroats}

In this section we will be interested in estimating the number of longer Klebanov-Strassler throat in a generic Type IIB Calabi-Yau flux compactification. In particular, we would like to get an idea about the maximum number of such longer throats. A statistical analysis for the number and length of such throats following the principles laid down in \cite{Douglas} was done in \cite{Hebecker:2006bn}.  Our discussion will be more general than this analysis.

Let us assume that the compact Calabi-Yau manifold has ${\sf N}_L=\sum_i {\sf N}_i$ number of large and $\sum_a n_a$ number of small Klebanov-Strassler throats. The radii of curvature $R_i$ of the former class of throats are large compared to the latter ones, $r_a$. We also assume that ${\sf N}_i \gg n_a$ as the larger throats have small curvatures and hence the classical supergravity analysis is valid.
\begin{figure}[!ht]
\begin{center}
\includegraphics[scale=0.45]{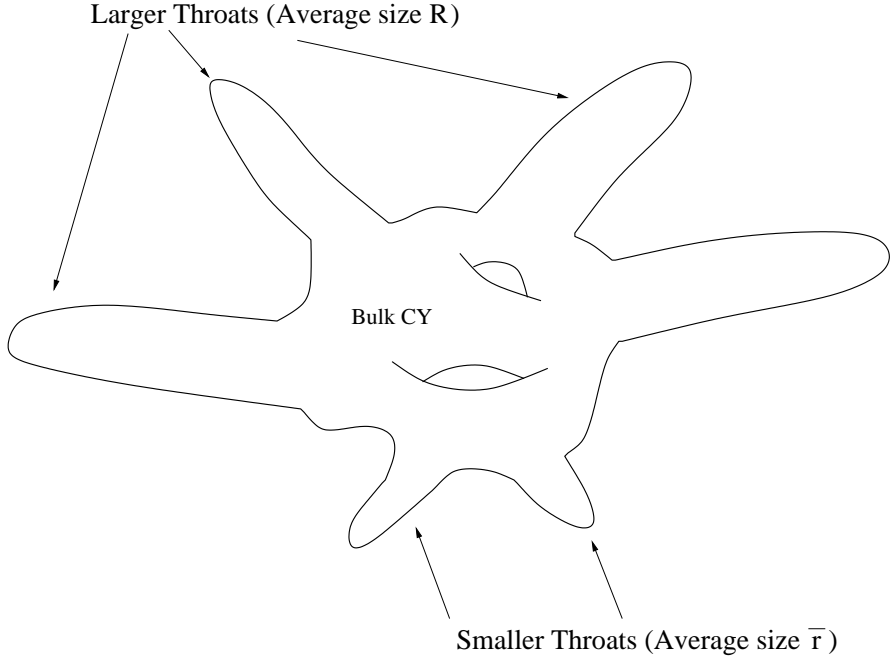}
\caption{A Calabi-Yau manifold with many larger throats and a few smaller throats. the average radius of curvature of the larger throats is $\bar{R}$. On an average they merge with the bulk Calabi-Yau at a distance $\bar{r}_{{\rm UV}}$, measured from their tip. The average radius of curvature of the smaller throats is $\bar{r}$.\label{fig2}}
\end{center}
\end{figure}
From \refb{FourDimPlanck}, \refb{CYVolume} and \refb{throatvolume}, it follows that
\be\label{largeandsmall}
M_P^2  = \frac{\rm{Vol}(\Omega_5)}{2g_s^2\kappa_{10}^2}
\left[\sum_i {\sf N}_i R_i^4 (r_{\rm{UV}})_i^2
+ \sum_a n_a r_a^4 (r_{\rm{UV}})_a^2 + \delta V^w_6\right] \ .
\ee
Here $(r_{\rm{UV}})_i$ and $(r_{\rm{UV}})_a$ are the distances of the larger and smaller throats, respectively, measured from the tip of the respective throats, where they merge smoothly with the rest of the Calabi-Yau manifold. From \refb{largeandsmall}, it follows that
\be
\frac{\rm{Vol}(\Omega_5)}{2g_s^2\kappa_{10}^2} \sum_i {\sf N}_i R_i^4 (r_{\rm{UV}})_i^2  < M_P^2 \ .
\ee
For estimation, let us assume that all the larger throats have almost same radii of curvature, $R_i\simeq\bar{R}$ (see Fig. \ref{fig2}). Moreover, we also assume that their merging points in the bulk of the Calabi-Yau are almost the same and equal to some average value $(r_{{\rm UV}})_i\simeq\bar{r}_{\rm{UV}}$. Since ${\sf N}_L = \sum_i {\sf N}_i$ is the total number of larger throats, we obtain
\be
 \frac{{\sf N} _L}{2g_s^2\kappa^2_{10}}{\rm{Vol}}(\Omega_5)\bar{R}^4\bar{r}_{{\rm UV}}^2 < M_P^2 \ .
\ee
Using \refb{adscurvature} and the relation $\bar{\phi}_{\rm{max}} = \sqrt{T_3}\bar{r}_{{\rm UV}}$, where $\bar{\phi}_{\rm{max}}$ is the average value of the maximum bound of the inflaton field inside the larger throats, we have
\be
{\sf N}_L < \left(\frac{M_P}{\bar{\phi}_{\rm{max}}}\right)^2 \frac{4}{\overline{M}\ \overline{K}} \ ,
\ee
where $\overline{M}$, $\overline{K}$ are some average values of the fluxes inside the larger throats.

The larger throats have smaller curvature, so in their presence we can trust the semi-classical supergravity analysis. Their presence makes the dynamics of the inflaton over the string theory landscape more interesting. In particular, one major dynamical problem is to find out the tunneling rates of the inflaton field between various larger Klebanov-Strassler throats. One can address this problem in the language of four-dimensional effective field theory. In general, the bulk Calabi-Yau geometry can be quite complicated and we expect that the classical traversing of the D3-brane from one throat to another may be impossible. In that case, we can look for the tunneling rate due to barrier penetration following references \cite{Coleman,CallanColeman,ColemanDeLuccia}.
%Generically the antisymmetric part of such tunneling rates may not vanish. This in turn will imply that the %probability dynamics of the inflaton on the landscape will be subjected to disorder.

%%%%%%%%%%%%%%%%%%%%%%%%%%%%%%%%%%%%%%%%%%%%%%%%%%%%%%%%%%%%%%%%%%%%%%%%%%%%%%%%%%%%%%%%
%%%%%%%%%%%%%%%%%%%%%%%%%%%%%%%%%%%%%%%%%%%%%%%%%%%%%%%%%%%%%%%%%%%%%%%%%%%%%%%%%%%%%%%%
%%%%%%%%%%%%%%%%%%%%%%%%%%%%%%%%%%%%%%%%%%%%%%%%%%%%%%%%%%%%%%%%%%%%%%%%%%%%%%%%%%%%%%%%
%%%%%%%%%%%%%%%%%%%%%%%%%%%%%%%%%%%%%%%%%%%%%%%%%%%%%%%%%%%%%%%%%%%%%%%%%%%%%%%%%%%%%%%%
%%%%%%%%%%%%%%%%%%%%%%%%%%%%%%%%%%%%%%%%%%%%%%%%%%%%%%%%%%%%%%%%%%%%%%%%%%%%%%%%%%%%%%%%

\section{Continuum limit of vacuum dynamics equations. Tunneling graph}\label{sec:continuum}

In the previous Section we have identified one source of disorder on the string theory landscape:
a sector on the landscape corresponding to a certain class of manifolds string theory can
be compactified to, namely, Calabi-Yau manifolds with a number of warped Klebanov-Strassler throats attached to the bulk part of the Calabi-Yau manifold (see Fig. \ref{fig2}). Since KS throats can be attached randomly to the bulk CY and have random lengths, dynamics of eternal inflation on the sector of the landscape will be affected by this randomness.

Let us now consider the vacuum dynamics equations (\ref{eq:VacuumDynamicsNonVolume}) on a \emph{generic} landscape (or a sector of the landscape).

To simplify the discussion, consider first a part of the landscape which consists only of de Sitter vacua, {\em i.e.}, vacua with positive cosmological constants. We want to understand the dynamics of tunneling between de Sitter vacua.
%We will then take into account the presence of AdS vacua
%(see discussion around equation \refb{eq:VacDynContinLimitVolumeSinks} below).

The vacuum dynamics equations (\ref{eq:VacuumDynamicsNonVolume})
define the dynamics of probabilities $P_{i}$ to measure a given value of the cosmological
constant in a given Hubble patch (\emph{i.e.}, for an observer to find herself
in a de Sitter vacuum $i$).
In general, the index $i$ enumerates
all possible de Sitter vacua on the landscape or just vacua within
a single (sub)sector of the landscape we are interested in. For example,
$i$ can label vacuum states with different number of fluxes such
as in the Bousso-Polchinski landscape \cite{PolchinskiBousso} or different
KS throats in the sector of the landscape corresponding to tunneling between
separate KS throats such as in the case discussed in Section \ref{sec:eternalKS}.

The tunneling rates between de Sitter vacua $i$ and $j$ are given by $\Gamma_{ij}$.
These tunneling rates are not necessarily given by the Coleman - de Luccia instantons: for a example, on the sector of the landscape considered in Section \ref{sec:eternalKS}, $\Gamma_{ij}$ are given by inverse characteristic times of $D3$-brane travel between different KS throats. Many other sectors on the string theory landscape may
exist due to a non-trivial dynamics of volume and complex structure
moduli of the CY manifold. In the following, we do not need to know the tunneling rates $\Gamma_{ij}$ explicitly. By studying the properties of the solutions of vacuum dynamics equations (\ref{eq:VacuumDynamicsNonVolume}) will let us draw lessons of physics interest which hold in general.
%Even without focusing on a particular form of tunneling rates $\Gamma_{ij}$, discussion
%of features of eternal inflation on the landscape in terms of the
%vacuum dynamics equations (\ref{eq:VacuumDynamicsNonVolume}) remains
%legitimate for an arbitrary sector of the landscape, and we can learn
%a lot about tunneling on the landscape just from general properties
%of solutions of the vacuum dynamics equations.

The general solution of the system (\ref{eq:VacuumDynamicsNonVolume})
can be represented as a sum \cite{LindeLandscapeLate}
%of exponentially decaying terms \cite{LindeLandscapeLate},
\be
 P_{i} (\tau) = \sum_{j=1}^{n_{\rm max}}A_{ij}e^{-\omega_{ij}\tau} \ ,
\ee
where $A_{ij}$ are constants of integration, $n_{\rm max}$ is the total number of vacua on the landscape and $\omega_{ij}$ are
complex (in the general case) functions of tunneling rates $\Gamma_{ij}$.
More precisely, the system of $n_{\rm max}$ first order differential equations
(\ref{eq:VacuumDynamicsNonVolume}) can be transformed to a single $n_{\rm max}^{\rm th}$ order
differential equation
\be
 \sum_{j=1}^{n_{\rm max}}a_{j}\frac{d^{j}P_{i}}{d\tau^{j}} = 0 \ . \label{eq:Pa}
\ee
In (\ref{eq:Pa}) $a_{j}$ are known functions of the tunneling rates $\Gamma_{ij}$. Then,
the rates $\omega_{ij}$ are given by $n_{\rm max}$ roots of the $n_{\rm max}^{\rm th}$ order
polynomial
\be
 \sum_{j=1}^{n_{\rm max}}a_{j}(-)^j \omega_{ik}^{j}=0 \ , \ \quad\forall\ i,k =1,\ldots,n_{{\rm max}}\ .\label{eq:RatesPolynomial}
\ee
Clearly, when the number of vacua $n_{\rm max}$ is very large,\footnote{On the ``multithroat'' sector of the landscape discussed in the Subsection \ref{sec:numthroats} it is not very large, but still we can take $n_{\rm max}$ of order of $100$.} finding the spectrum of $\omega_{ij}$'s by directly solving the polynomial equation
(\ref{eq:RatesPolynomial}) is hopeless. However, when the tunneling rates $\Gamma_{ij}$ are weakly fluctuating functions of the position $i$ on the landscape (and $n_{\rm max}\gg 1$) the vacuum dynamics equations (\ref{eq:VacuumDynamicsNonVolume}) have a continuum limit. In this limit we can adopt powerful methods of the theory of partial differential equations and functional analysis.
% \lan{} and one needs to introduce
%another strategy to solve the system (\ref{eq:VacuumDynamicsNonVolume}).
%When tunneling rates $\Gamma_{ij}$ are weakly fluctuating functions
%of the position $i$ on the landscape and $n_{\rm max}\gg1$, one such strategy
%is to analyze the vacuum dynamics equations in the continuous
%limit, where powerful methods of PDE and functional analysis are applicable.

\begin{figure}
\begin{center}
\includegraphics[scale=0.9]{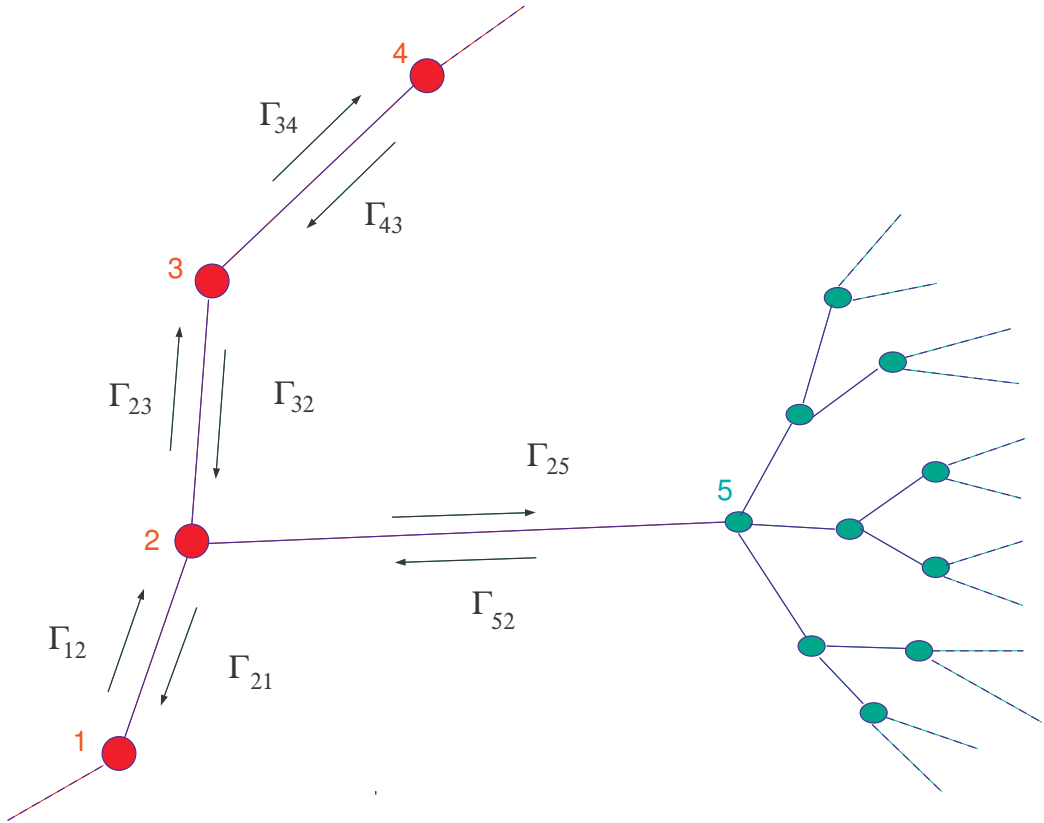}
\caption{The tunneling graph of vacua. The island of vacua denoted by green
is well separated from ``red'' vacua because the tunneling rates
$\Gamma_{25}$, $\Gamma_{52}$ are small. The dimensionality $N$
of the island of ``red'' vacua is $1$ (two nearest neighbors
exist for any given ``red'' vacuum), while $N=2$ for the island
of ``green'' vacua.\label{fig:Vacua}}
\end{center}
\end{figure}

%In the present Section our goal is to understand when the continuous
%limit of the Eqs. (\ref{eq:VacuumDynamicsNonVolume}) exists.
To determine how the continuum limit of the vacuum dynamics equations
(\ref{eq:VacuumDynamicsNonVolume}) emerges, we notice
that (\ref{eq:VacuumDynamicsNonVolume}) define dynamics
of tunneling (hopping) on a graph (see Fig. \ref{fig:Vacua}):
each vacuum is a node in the graph denoted by the (composite) index
$i$, and links between different nodes have weights given by tunneling
rates $\Gamma_{ij}$. As we have depicted in Fig. \ref{fig:Vacua}, parts of the graph can be isolated. These are called islands on the landscape (see also \cite{LindeLandscapeEarly,LindeLandscapeLate}). The Hausdorff dimension $N$ of a particular island is related to the number of nearest neighbors in such a tunneling graph. For example, for vacua represented by red dots in Fig. \ref{fig:Vacua}, $N = 1$. On the other hand, $N=2$ for the vacua represented by green dots in Fig. \ref{fig:Vacua}.
%The Hausdorff dimension $N$ of the tunneling graph
%or a part of the graph which we denote as an island on the
%landscape (see also \cite{LindeLandscapeEarly,LindeLandscapeLate})
%is defined as a number of nearest neighbors to any given vacuum within
%the island.

For simplicity, we will first consider the quasi-one-dimensional case $N=1$ (two nearest neighbors for any given vacuum on the island). In this case,
exactly two adjacent vacua exist for any given vacuum $i$ on the island and tunneling
rates to other vacua from $i$ are suppressed. Following the terminology
introduced in \cite{PodolskyEnqvist} such islands are denoted as quasi-one-dimensional
islands on the landscape. The system of equations (\ref{eq:VacuumDynamicsNonVolume})
 reduces to
\be
 \frac{dP_{i}}{d\tau} = \Gamma_{i+1,i}P_{i+1}-\Gamma_{i,i+1}P_{i}+\Gamma_{i-1,i}P_{i-1}-\Gamma_{i,i-1}P_{i} \ . \label{eq:VacuumDynamicsNonVolume1D}
\ee

First of all, let us suppose that tunneling rates $\Gamma_{i,i+1}$
are slightly fluctuating near some average value $\bar{\Gamma}$, so
that they can be represented as
\be
 \Gamma_{i,i+1}=\bar{\Gamma}+\delta\Gamma_{i,i+1}^{s}+\delta\Gamma_{i,i+1}^{a}\ ,
\ee
where
\be
 \delta\Gamma_{i,i+1}^{s}=\frac{1}{2}\left(\delta\Gamma_{i,i+1}+\delta\Gamma_{i+1,i}\right)
\ee
is symmetric part of the tunneling rate and
\be
 \delta\Gamma_{i,i+1}^{a}=\frac{1}{2}\left(\delta\Gamma_{i,i+1}-\delta\Gamma_{i+1,i}\right)
\ee
is its antisymmetric part. Apparently, this situation can be realized
in the multi-throat scenario we have discussed in Section \ref{sec:eternalKS}
-- for this, it is necessary that the lengths of different KS throats
and the distances between them in the interior CY were not very much
different. %\ack{Can we make this more precise?}

Let us rewrite the system (\ref{eq:VacuumDynamicsNonVolume1D}) as
\be
 \frac{dP_{i}}{d\tau}=\bar{\Gamma}\left(P_{i+1}-2P_{i}+P_{i-1}\right)+\Delta \ ,
\label{VDlatticized}
\ee
where
\bea
 \Delta & = & \delta\Gamma_{i,i+1}^{s}(P_{i+1}-P_{i})-\delta\Gamma_{i,i-1}^{s}(P_{i}-P_{i-1})+ \\
        &   & +\delta\Gamma_{i+1,i}^{a}(P_{i+1}+P_{i})-\delta\Gamma_{i,i-1}^{a}(P_{i}+P_{i-1}) \ .
\eea
%We conclude that the continuous limit of (\ref{eq:VacuumDynamicsNonVolume1D})
%exists if tunneling rates do not fluctuate too strongly for different $i$.
It is apparent from (\ref{VDlatticized}) that the continuum limit of (\ref{eq:VacuumDynamicsNonVolume1D}) exists and it is realized as a diffusion equation
\be
 \frac{\partial P}{\partial\tau}=D\frac{\partial^{2}P}{\partial n^{2}} \quad ; \quad D=\bar{\Gamma}\delta n^{2} \ , \label{eq:DiffsionEquation1DSimpified}
\ee
when $\Delta$ is set to zero. For non-vanishing $\Delta\ne 0$, (\ref{eq:DiffsionEquation1DSimpified}) has additional terms and it reads
\be
 \frac{\partial P}{\partial\tau}=\frac{\partial}{\partial n}\left((D+\delta D^{s}(n))\frac{\partial P}{\partial n}\right)+2\frac{\partial}{\partial n}\left(\delta D^{a}(n)P\right) \ , \label{eq:DiffusionEquation1D}
\ee
where $\delta D^{s}=\delta\Gamma^{s}\delta n^{2}$, $\delta D^{a}=\delta\Gamma^{a}\delta n^{2}$
and all higher order terms with respect to $\delta n$ are neglected.

The physical meanings of symmetric and antisymmetric fluctuating parts $\delta \Gamma_{ij}^s,\delta \Gamma_{ij}^a$
of the tunneling rate are clear. Equation (\ref{eq:VacuumDynamicsNonVolume1D})
describes a Brownian motion of a single ``particle'' in a
random medium.\footnote{Technically, this means that the motion of the particle is
determined by the Langevin equation $\frac{\partial n}{\partial\tau}=\eta(n,\tau)+F(n)$, where $\eta$ is a random force with correlation properties of a white noise $\langle\eta(n,\tau)\eta(n,\tau')\rangle = 2(D+\delta D^{s}(n))\delta(\tau-\tau')$. Equation (\ref{eq:DiffusionEquation1D}) is the corresponding Fokker-Planck equation.}
Symmetric part of the tunneling rate $\delta\Gamma_{ij}^{s}$ corresponds
to a fluctuation of the diffusion coefficient $D$, while antisymmetric
part $\delta\Gamma_{ij}^a$ to a random force acting on the particle while propagating in the medium. Since symmetric and antisymmetric fluctuation parts are independent we can separately analyze the ``random force'' and effects associated to the fluctuations of the diffusion coefficient $D$.

Let us now include AdS sinks to the discussion %. Let us
and also take into account the effects coming from finite volumes of expanding de Sitter bubbles. To understand the dynamics of the volume-weighted
probability \cite{LindeLandscapeEarly,LindeLandscapeLate,LindeMeasure}
we need to add extra terms $3H_{i}P_{i}$ (if $\tau$ is the world time of the four-dimensional observer) to the right-hand side of (\ref{eq:VacuumDynamicsNonVolume}). The continuum limit of
the vacuum dynamics equations for a quasi-one-dimensional island on
the landscape with AdS sinks then takes the form
\be
 \frac{\partial P}{\partial\tau}=\frac{\partial}{\partial n}\left((D+\delta D^{s}(n))\frac{\partial P}{\partial n}\right)+2\frac{\partial}{\partial n}\left(\delta D^{a}(n)P\right)+(3H(n)-\Gamma^s(n))P \ , \label{eq:VacDynContinLimitVolumeSinks}
\ee
where $\Gamma^s(n)$ is the rate of tunneling from a given vacuum $n$ to an AdS sink.

By performing similar analysis for quasi-2d ($N=2$), quasi-3d ($N=3$) etc. islands on the
landscape one finds the continuum limit of the vacuum dynamics equations
\be
 \frac{\partial P}{\partial\tau}=D\triangle P+\sum_{d}^N\left(\frac{\partial}{\partial n_{d}}\left(\delta D_{d}^{s}\frac{\partial P}{\partial n_{d}}\right)+2\frac{\partial}{\partial n_{d}}(\delta D_{d}^{a}P)\right) \ \label{eq:DiffusionEquationMultiD}
\ee
if AdS sinks and the volume effects of dS bubbles are neglected and
\be
 \frac{\partial P}{\partial\tau}=D\triangle P+\sum_{d}^N\left(\frac{\partial}{\partial n_{d}}\left(\delta D_{d}^{s}\frac{\partial P}{\partial n_{d}}\right)+2\frac{\partial}{\partial n_{d}}(\delta D_{d}^{a}P)\right)+(3H-\Gamma_{s})P
\ee
in the general case. The main difference from the quasi-one-dimensional
case ($N=1$) is that the diffusion coefficients $\delta D_{d}^{s}$ and $\delta D_{d}^{a}$
are now vectors; the diffusion of the ``particle''
is anisotropic (with $\delta D_{d}^{s}$ being the fluctuating anisotropic
diffusion coefficient), while the random force $\delta D_{d}^{a}$
acting on the ``particle'' is a vector. Effects due to the fluctuating
diffusion coefficient and random force will be treated separately
in the next two sections.

As a warm up before studying effects of randomness on the landscape,
let us briefly discuss continuum limit of Bousso-Polchinski (BP)
landscape \cite{PolchinskiBousso}. The BP landscape consists of a
discrete set of dS vacua with different values of the cosmological
constant $\Lambda$ defined by the formula
\be
 \Lambda=-\Lambda_{0}+\frac{1}{2}\sum_{k}q_{k}^{2}n_{k}^{2} \ ,
\ee
where $\Lambda_0$ is a bare vacuum energy, $n_{k}$ are the numbers of units of a given flux $k$, and $q_{k}$
are the associated topological charges.

The tunneling rates between different dS vacua are written in terms of Coleman - De Luccia (CDL) instantons \cite{ColemanDeLuccia,Parke,BrownTeitelboim,SchwartzPerlovVilenkin,CliftonShenkerSivanandam}
\be
 \Gamma_{ij}=A_{ij}e^{-B_{ij}} \ ,
\ee
where $A_{ij}\sim1$ and $B_{ij}$ are corresponding CDL actions. For adjacent vacua the action $B_{i,i-1}$ can be estimated
\cite{SchwartzPerlovVilenkin} as\footnote{For simplicity we consider flat spacetime limit.}
\be
 B_{i,i-1}\simeq \frac{27\pi^{2}}{8}\frac{1}{n_{i}^{3}q_{i}^{2}} \ . \label{eq:SchwartzPerlovVilenkin}
\ee
The formula (\ref{eq:SchwartzPerlovVilenkin}) is only valid when the tunneling exponent is large, \emph{i.e.}, at
small energies, when the values of cosmological constant in adjacent
vacua are relatively small compared to the Planckian density \cite{SchwartzPerlovVilenkin}.
%\be
%B_{ij} = B_{i\to j}\sim\exp\left(-\frac{24\pi^{2}}{\Lambda_{i}}+\frac{24\pi^{2}}{\Lambda_{j}}\right) \ . %\label{eq:ColemanDeLuccia}
%\ee
As we can see from (\ref{eq:SchwartzPerlovVilenkin}), the transitions between adjacent vacua happen faster at higher $n_i$, while
at low energies they proceed slower. Another important feature of
the BP landscape is that the transition downwards (towards lower energies)
is always more probable than the transition upwards -- from lower
to higher energies, because tunneling rates for adjacent vacua are related as
\be
 \Gamma_{i-1,i} = \Gamma_{i,i-1} \exp \left( S_{i} - S_{i-1} \right) = \Gamma_{i,i-1} \exp ( \delta S_i ),
\ee
where
\be
S_i \sim 1/\Lambda_i
\ee
is the Gibbons-Hawking entropy.

Let us consider transitions between vacua different by the number
of quanta of a single flux $n_{i}\to n_{i}-1$. For the symmetric
and antisymmetric parts of the tunneling rate $\Gamma_{ij}$ between
adjacent dS vacua we have for $q_{i}n_{i}\ll\Lambda_{i}$,
\be
 \Gamma_{i,i-1}^{s} = \frac{1}{2}(\Gamma_{i,i-1}+\Gamma_{i-1,i})\approx A\cdot \exp \left( - B_{i,i-1}\right)\cdot
 \left(1 + \frac{1}{2}\delta S_i \right)
\ee
and
\be
\Gamma_{i,i-1}^{a}=\frac{1}{2}(\Gamma_{i,i-1}-
\Gamma_{i-1,i})\approx A \exp \left( - B_{i,i-1}\right)\cdot
  \frac{1}{2}\delta S_i  ,
\ee
where we took $\delta S_i \ll 1$.

Therefore, the continuum version of the vacuum dynamics equations
for this sub-sector of the BP landscape (tunneling between vacuum states
different only by the number of quanta of a single flux) is
\be
 \frac{\partial P}{\partial\tau}=A\frac{\partial}{\partial n}\left(  A\cdot e^{-B_{i,i-1}}\cdot
 \left(1 + \frac{1}{2}\delta S_i \right) \frac{\partial P}{\partial n}\right)+A\frac{\partial}{\partial n} \left( e^{-B_{i,i-1}}\cdot \delta S_i \right) \ .
 \label{eq:FPBoussoPolchinski}
\ee
This equation describes diffusion under the action of external ``force''
$F=Ae^{-B_{i,i-1}}\delta S_i$ directed from higher
$n_{i}$ (vacua with larger values of the cosmological constant) towards
lower $n_{i}$ (vacua with lower values of the cosmological constant).

Equation (\ref{eq:FPBoussoPolchinski}) describes a single quasi-one-dimensional ($N=1$) sector of the BP landscape.
It generalizes straightforwardly to other possible transition channels between vacua with different fluxes turned on. Entire BP landscape is described by a similar equation with $N\gg 1$ and all possible transition channels taken into account.

%The generalization of the Eq. (\ref{eq:FPBoussoPolchinski}) to other
%possible channels of transitions between vacua of the BP landscape
%is straightforward. \lan{Sounds good?}

%%%%%%%%%%%%%%%%%%%%%%%%%%%%%%%%%%%%%%%%%%%%%%%%%%%%%%%%%%%%%%%%%%%%%%%%%%%%%%%%%%%%%%%%
%%%%%%%%%%%%%%%%%%%%%%%%%%%%%%%%%%%%%%%%%%%%%%%%%%%%%%%%%%%%%%%%%%%%%%%%%%%%%%%%%%%%%%%%
%%%%%%%%%%%%%%%%%%%%%%%%%%%%%%%%%%%%%%%%%%%%%%%%%%%%%%%%%%%%%%%%%%%%%%%%%%%%%%%%%%%%%%%%
%%%%%%%%%%%%%%%%%%%%%%%%%%%%%%%%%%%%%%%%%%%%%%%%%%%%%%%%%%%%%%%%%%%%%%%%%%%%%%%%%%%%%%%%
%%%%%%%%%%%%%%%%%%%%%%%%%%%%%%%%%%%%%%%%%%%%%%%%%%%%%%%%%%%%%%%%%%%%%%%%%%%%%%%%%%%%%%%%

\section{Effects of disorder on the landscape}\label{sec:Disorder}

In this Section we analyze effects of disorder on the landscape. In particular, we study solutions of the vacuum dynamics equations for the non-volume-weighted measure with disorder built in the tunneling rates $\Gamma_{ij}$.
We are especially interested to find how this disorder affects long time (large but finite $\tau$) behavior of the probability distribution $P(\vec{n},\tau )$
for a $4$-dimensional observer to measure a given value of the cosmological constant $\Lambda_n$ on the sector of the landscape discussed in the Section \ref{sec:eternalKS}.
To determine this behavior, we pursue the following strategy. First, we suppose that tunneling rates $\Gamma_{ij}$ are not too strongly fluctuating functions of the position on the landscape (index $i$ labeling vacua on the island);
since the number of vacua on the landscape (or an island on the landscape) is considered to be very large, we use continuous version of the vacuum dynamics equations (\ref{eq:DiffusionEquationMultiD}) derived in Section \ref{sec:continuum}. Second, we study asymptotic properties of the general solution of the vacuum dynamics equations by averaging over disorder present in tunneling rates $\Gamma_{ij}$ and by using dynamical renormalization group methods.

\subsection{Random environment}\label{sec:Random-environment}

We begin analyzing (\ref{eq:DiffusionEquation1D})
and (\ref{eq:DiffusionEquationMultiD}) by studying influence of the
``random environment'' (fluctuating antisymmetric parts of tunneling
rates $\Gamma_{ij}^{a}$) on the dynamics of the probability distribution
$P(\vec{n},\tau)$ in the presence of disorder on the landscape. Namely,
we consider the $N$-dimensional diffusion equation
\be
\frac{\partial P}{\partial\tau}=D\triangle P+2\sum_{d=1}^{N}\frac{\partial}{\partial n_{d}}(\delta D_{d}^{a}P)\ , \label{eq:DiffusionEquationRandomForce}
\ee
where the ``random force'' $\delta D_{d}^{a}$ has the correlation
properties
\bea
\langle\delta D_{d_{1}}^{a}(\vec{n})\delta D_{d_{2}}^{a}(\vec{n}')\rangle & = & \sigma\delta_{d_{1},d_{2}}\delta(\vec{n}-\vec{n}') \label{eq:RandomForceGaussian} \\
\langle\delta \vec D^{a}(\vec{n})\rangle & = & 0 \label{eq:RandomForceGaussian2}
\eea
of the white noise. Here $\sigma$ is a constant denoting disorder strength.

Such correlation properties correspond to a multithroat sector of the landscape discussed in the Section \ref{sec:eternalKS}. In particular, the absence of any disorder corresponds to the situation when KS throats have equal lengths and are attached equidistantly to the bulk CY. Weak disorder corresponds to slight fluctuations of their lengths and distances between them.

Strictly speaking, condition (\ref{eq:RandomForceGaussian2}) is not applicable, for example, for the BP landscape, where transitions towards smaller $n$ are always more probable. This gives rise to the correlation properties
\be
\langle\delta \vec D^{a}(\vec{n})\rangle \ne 0
\ee
to be discussed in the future publication.

We study properties of the solution of (\ref{eq:DiffusionEquationRandomForce}) using methods of dynamical renormalization group analysis.\footnote{Non-gaussian correlation properties different from (\ref{eq:RandomForceGaussian}) give rise to irrelevant terms in the effective action (see (\ref{eq:RandomForceEffectiveAction})
below) and do not affect late time behavior of the probability
distribution $P(\vec{n},\tau)$.}

Equation (\ref{eq:DiffusionEquationRandomForce}) (with correlation
properties (\ref{eq:RandomForceGaussian}),(\ref{eq:RandomForceGaussian2}) of $\delta \vec D^{a}$) can
be considered as a Fokker-Planck equation describing a random walk of
a particle in a random environment (characterized by the coefficients $\delta D_d^{a}(\vec{n})$).
Indeed, if the particle is moving according to the Langevin equation
\be
 \frac{\partial\vec n(\tau)}{\partial\tau} = \vec\eta(\tau) + \delta \vec D^{a}(\vec{n}) \ ,
\ee
where the noise $\vec{\eta}(\tau)$ is correlated as
\bea
 \langle\eta^{b_{1}}(\tau)\eta^{b_{2}}(\tau')\rangle & = & D\delta^{b_{1},b_{2}}\delta(\tau-\tau') \\
 \langle\vec{\eta}\rangle & = & 0 \ ,
\eea
then (\ref{eq:DiffusionEquationRandomForce}) defines the probability
to find the particle in the position $\vec{n}$ at a given time $\tau$.
Apart from the random noise $\vec{\eta}$ the particle is affected
by the force $\vec{\delta D^{a}}$ in any point of the medium.

%\lan{Good below?} We will follow the analysis of random walks in random environments
%by Daniel Fisher \cite{Fisher}.
To analyze random walks in a random environment, we will closely follow \cite{Fisher}.
Let us hence introduce the Martin-Siggia-Rose (MSR) generating functional \cite{MartinSiggiaRose}
\bea
Z[J] & = & \int{\cal D}P{\cal D}\delta\vec{D^{a}}\ \delta\left(\frac{\partial P}{\partial\tau}-D\triangle P-2\sum_{d}\frac{\partial}{\partial n_{d}}(\delta D_{d}^{a}P)\right) \nonumber\\
 & & \times\exp\left(\int d\tau d^{N}n J(\tau,\vec{n})P(\tau,\vec{n})\right)\exp\left(-\int d^{N}n\frac{{(\delta\vec D^{a})^{2}}}{\sigma}\right) \
\eea
and average over disorder present in $\delta D_{d}^{a}$.

By introducing Lagrangian multipliers and transforming the functional $\delta$-function into the exponent,
the latter can be rewritten in terms of MSR field variables $\bar{P}$
and $P$ \cite{SchwingerKeldysh,MartinSiggiaRose}
\bea
 Z[J] & = & \int{\cal D}\bar{P}{\cal D}P{\cal D}\delta\vec D^{a}\exp(S_{{\rm MSR}}^{a}[\bar{P},\, P,\,\delta\vec D^{a}]) \nonumber\\
      & & \times\exp\left(\int d\tau\, d^{N}n\, J(\tau,\vec{n})\, P(\tau,\vec{n})\right)\exp\left(-\int d^{N}n\frac{(\delta\vec D^{a})^2}{\sigma}\right) \ ,
\eea
where
\be
 S_{{\rm MSR}}^{a}[\bar{P},\, P,\,\delta\vec D^{a}] = i\int d\tau d^{N}n\,\bar{P}\left(\frac{\partial P}{\partial\tau}-D\triangle P-2\sum_{d=1}^N\frac{\partial}{\partial n_{d}}(D_{d}^{a}P)\right) \ .
\ee
Integrating out the Gaussian disorder $\delta D^a$ we find
\be
Z[J]=\int{\cal D}\bar{P}{\cal D}P\,\exp(S_{{\rm eff}}^{a}[\bar{P},\, P])\exp\left(\int d\tau d^{N}n\, J(\tau,\vec{n})P(\tau,\vec{n})\right) \ ,
\ee
where the effective action reads
\bea\label{eq:EffAction}
S_{{\rm eff}}^{a} & = & i\int d\tau d^{N}n\,\bar{P}\left(\frac{\partial P}{\partial\tau}-D\triangle P\right) \label{eq:RandomForceEffectiveAction}\\
 & & -\frac{\sigma}{2}\sum_{d=1}^N\int d\tau d\tau'd^{N}n\, P(\vec{n},\tau)\frac{\partial}{\partial n_d}\bar{P}(\vec{n},\tau)P(\vec{n},\tau')\frac{\partial}{\partial n_d}\bar{P}(\vec{n},\tau') \ . \nonumber
\eea
The bare propagator of the field $P$ in the momentum representation is
\be
 G_{0}(\omega,q)=\frac{1}{-i\omega+Dq^{2}} \ ,
\ee
where $\omega$ and $\vec{q}$ are the canonically conjugate momentum variables of
$\tau$ and $\vec{n}$, respectively. Hence in the absence of disorder, the probability distribution $P(\vec{n},\tau)$ is spreading
out according to the linear diffusion law
\be
 \langle\vec{n}^{2}(\tau)\rangle =  2 N D\tau . \label{eq:LinearDiffusionLaw}
\ee
%independent of the effective dimensionality $N$.

In order to find the late time behavior of $P(\vec{n},\tau)$, one should, in principle, work it out from \refb{eq:EffAction} perturbatively in $\sigma$ using standard diagrammatic techniques. However, using simple scaling arguments we can determine the late time
behavior of the probability distribution $P(\vec{n},\tau)$. From the quadratic part of the action, by multiplying momentum $\vec{q}$ by a factor of $l$, we immediately find that the effective frequency
$\omega$ rescales as $\omega'=\omega l^{z}=\omega l^{2}$, while
the Fourier components of the MSR fields $\widetilde{P}(\omega,\vec{q})$ and
$\widetilde{\bar{P}}(\omega,\vec{q})$ rescale as $\widetilde{P}\zeta$ and
$\widetilde{\bar{P}}\bar{\zeta}$,
where $\zeta\bar\zeta=l^{N+4}$. The same argumentation for the interaction term proportional to $\sigma$
in the effective action (\ref{eq:RandomForceEffectiveAction}), shows that the disorder
strength $\sigma$ renormalizes as $\sigma'=l^{2-N}\sigma$. Therefore,
disorder should be irrelevant for islands on the landscape with $N>2$, and the linear diffusion
law (\ref{eq:LinearDiffusionLaw}) is not affected by disorder at long time scales.
%$\tau\to\infty$.

\begin{figure}
\includegraphics[scale=1.2]{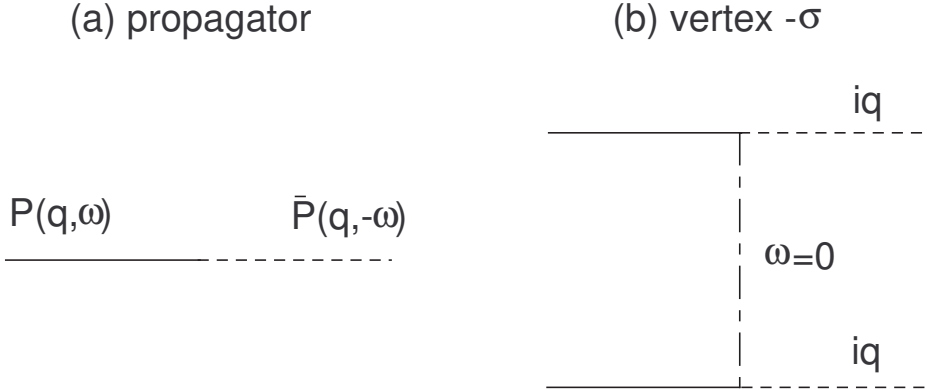}
\caption{The MSR propagator and the vertex proportional to the strength of
the disorder.\label{fig:propa}}
\end{figure}

To understand what happens at $N\leq2$ we have to generalize these
naive scaling arguments into complete renormalization group analysis
(expansion in the disorder strength $\sigma$ near
$N=2$). Diagrammatic rules following from the form of the MSR action
(\ref{eq:RandomForceEffectiveAction}) are given by the bare propagator
$iG_{0}(\omega,q)=(\omega-iDq^{2})^{-1}$ and the vertex $\sim-\sigma/2$
represented in Fig. \ref{fig:propa}.

At one-loop level the propagator remains unchanged \cite{Fisher}, but the vertex renormalizes as
\be
 \frac{d\sigma}{dl} = (2-N)\sigma-\frac{1}{4\pi}\sigma^{2} + \mathcal O(\sigma^{3}) \ . \label{eq:RenormalizationSigma}
\ee
The RG flow for the dynamical exponent $z$ is given by
\be
 \frac{dz}{dl}=2+\frac{1}{8\pi^{2}}\sigma^{2}+\mathcal O(\sigma^{3}) \ . \label{eq:RenormalizationZ}
\ee
%In the absence of disorder
$\sigma=0$, $z=2$ is a fixed point of the renormalization group flow. A second-order
analysis shows that it is the only fixed point for $N>2$ \cite{Fisher}. Therefore,
disorder does not affect the late time asymptotics of the probability
distribution $P(\vec{n},\tau)$ for large $N$, and the mean-square
displacement $\langle\vec{n}^{2}(\tau)\rangle$ behaves according
to the linear diffusion law (\ref{eq:LinearDiffusionLaw}) at large $\tau$. %$\tau=\infty$.

At $N=2-\epsilon$, long time evolution is determined by a new fixed
point
\bea
\sigma & = & 4\pi\epsilon + \mathcal O(\epsilon^{2}) \label{eq:SigmaNonTrivialFixedPoint} \\
 z     & = & 2+2\epsilon^{2} + \mathcal O(\epsilon^{3})\ . \label{eq:ZNonTrivialFixedPoint}
\eea
The mean-square displacement
\be
 \langle\vec{n}^{2}(\tau)\rangle = -\frac{d}{dq^d}\frac{d}{dq^d}\int d\omega e^{-i\omega\tau}G(\omega,\vec{q})|_{\vec{q}=0}
\ee
behaves as
\be
 \langle\vec{n}^{2}(\tau)\rangle\sim\tau^{2/z} \sim \tau^{1-\epsilon^{2}} \ . \label{eq:n2epsilon}
\ee
For $N<2$ the diffusion process is slow, and $\epsilon=1$ (quasi-one-dimensional
islands on the landscape) is a special case treated separately below.

%Before turning to the $N=1$, let us discuss the special case $N=2$
%in more details. From the Eq. (\ref{eq:RenormalizationSigma}) one
%concludes that the strength of disorder $\sigma$ is renormalized
%as\[
%\frac{d\sigma}{dl}=-\frac{1}{4\pi}\sigma^{2}+O(\sigma^{3}),\]
%\emph{i.e.}, \[
%\sigma(l)=\frac{\sigma_{0}}{1+\frac{\sigma_{0}l}{4\pi}}.\]
%Inverse characteristic time scale $\omega=t^{-1}$ is renormalized
%according to\[
%\frac{d\omega}{dl}=z\omega=\left(2+\frac{\sigma^{2}}{8\pi^{2}}\right)\omega,\]
% \[
%\log\left(\frac{\omega(l)}{\omega_{0}}\right)=2l-\frac{1}{2\pi}\left(\frac{\sigma_{0}}{1+\frac{\sigma_{0}l}{4\pi}}-\s5igma_{0}\right).\]

Similarly, one can show that for $N=2$ the mean-square displacement
behaves as
\be
 \langle\vec{n}^{2}(\tau)\rangle = 4D_{R}\tau\left(1+\frac{4}{\log\tau}+\ldots\right) \ ,
\ee
where $D_{R}$ is the renormalized diffusion coefficient, \emph{i.e.}, diffusion
is linear with a small logarithmic correction.

%\lan{Good below?}
%The long time behavior of the distribution function for $N>2$ is determined at weak disorder by
%the trivial fixed point $\sigma=0$, and diffusion proceeds according
%to the usual linear law as in the absence of any disorder.
For $N>2$ and at weak disorder $\sigma\simeq 0$, the long time behavior is determined by the trivial fixed point $\sigma=0$ corresponding to the usual linear diffusion law.
%Strictly at $\sigma=0$, the diffusion proceeds according to the usual linear diffusion law.

It is interesting to note that the system of islands on the landscape
in fact introduces a \emph{physical meaning} to the RG $\epsilon$-expansion.
This is because the number of nearest neighbors for a
vacuum in a given island can fluctuate within the island. The
Hausdorff dimensionality of the island (see Section \ref{sec:continuum})
can easily have non-integer values if the tunneling graph of the
island is fractal-like. In this case, the long time behavior of the
mean-square displacement is given by
\be
 \langle\vec{n}^{2}(\tau)\rangle\sim\tau^{1-(2-N)^{2}} = \tau^{-3+4N-N^{2}}
\ee
for $1<N<2$, and the diffusion of the probability distribution is
anomalous.

The case $N=1$ turns out to be the most interesting. If we naively substituted
$\epsilon=1$ into the formula (\ref{eq:n2epsilon}), we would expect the diffusion
process to be frozen even for weak disorder. This turns out to be too premature. Indeed, Sinai
showed by using ergodic methods \cite{Sinai}, that the diffusion process does not completely
stop but proceeds instead very slowly according to the logarithmic
law
\be
 \langle n^{2}(\tau)\rangle \sim \log^{4}\tau \ .
\ee
Notice that this behavior is independent of the disorder strength (the
result holds for an arbitrarily weak disorder) and the disorder potential.\footnote{Non-exponential behavior of $P_i(\tau)$ was also reported in \cite{Winitzki:2007cf} in the different context (there the situation was considered when transition probability between vacua depends on the age of each vacuum). }

The overall conclusion we come to in this Section is the following.
For islands on the landscape with $N<2$ of the tunneling graph, the diffusion of the probability
to measure a given value of the cosmological constant in a given Hubble
patch is suppressed according to the formula (\ref{eq:n2epsilon})
for any amount of the ``environment'' disorder $\sigma$. More precisely,
it is suppressed logarithmically for quasi-one-dimensional islands ($N=1$)
on the landscape and as a slow power law for $1<N<2$. For quasi-two-dimensional
islands ($N=2$),  the linear diffusion law acquires small logarithmic corrections,
and for $N>2$ the diffusion of the probability distribution function $P$ evolves
according to the linear diffusion law.

%%%%%%%%%%%%%%%%%%%%%%%%%%%%%%%%%%%%%%%%%%%%%%%%%%%%%%%%%%%%%%%%%%%%%%%%%%%%%%%%%%%%%%%%
%%%%%%%%%%%%%%%%%%%%%%%%%%%%%%%%%%%%%%%%%%%%%%%%%%%%%%%%%%%%%%%%%%%%%%%%%%%%%%%%%%%%%%%%
%%%%%%%%%%%%%%%%%%%%%%%%%%%%%%%%%%%%%%%%%%%%%%%%%%%%%%%%%%%%%%%%%%%%%%%%%%%%%%%%%%%%%%%%
%%%%%%%%%%%%%%%%%%%%%%%%%%%%%%%%%%%%%%%%%%%%%%%%%%%%%%%%%%%%%%%%%%%%%%%%%%%%%%%%%%%%%%%%
%%%%%%%%%%%%%%%%%%%%%%%%%%%%%%%%%%%%%%%%%%%%%%%%%%%%%%%%%%%%%%%%%%%%%%%%%%%%%%%%%%%%%%%%

\subsection{Random anisotropic diffusion}\label{sec:Anisotropic-diffusion}

We now turn to the discussion of effects due to fluctuations of the
anisotropic diffusion coefficient $\delta \vec D^s$. Consider the following diffusion
equation
\be\label{eq:aniso}
\frac{\partial P}{\partial\tau}=\sum_{d}^N\frac{\partial}{\partial n^d}\left((D+\delta D_d^s(n))\frac{\partial P}{\partial n^d}\right) \ ,
\ee
where fluctuating anisotropic diffusion coefficients $\delta D^s_d$ have the correlation
properties\footnote{We again consider only the multithroat sector on the landscape described in the Section \ref{sec:eternalKS}. Correlation properties of $D_{\alpha}^{s}$ are different for the BP landscape ($\langle\delta D_{\alpha}^{s}(\vec{n})\rangle \ne 0$) and will be considered elsewhere.}
\bea
\langle\delta D_{d_1}^{s}(\vec{n})\delta D_{d_2}^{s}(\vec{n}')\rangle & = & \Delta\delta_{d_1,d_2}\delta(\vec{n}-\vec{n}') \label{eq:RandomDiffusionCorrProperties} \\
\langle\delta D_{\alpha}^{s}(\vec{n})\rangle & = & 0 \ .
\eea
%where $\Delta$ is the diffusion coefficient.

In principle, it is necessary to consider only the case $\delta D^{s}<D$,
so these correlation properties should be modified. However, in the
renormalization group analysis this modification will introduce higher
order (\emph{i.e.}, irrelevant) terms in the effective action, and we will
neglect them in the present paper.

The Martin-Siggia-Rose generating functional now has the form
\bea
 Z[J] & = & \int{\cal D}P{\cal D}\delta\vec D^a\delta\left(\frac{\partial P}{\partial\tau}-\sum_{d}^N\frac{\partial}{\partial n^d}\left((D+\delta D_d^s(n))\frac{\partial P}{\partial n^d}\right)\right) \\
 & & \times\exp\left(\int d\tau d^{N}n\, J(\tau,\vec{n})P(\tau,\vec{n})\right)\exp\left(-\int d^{N}n\frac{\vec{(\delta D^{s})^{2}}}{\Delta}\right) \nonumber\\
  & = & \int{\cal D}\bar{P}{\cal D}P{\cal D}\delta\vec D^s\exp(S_{{\rm MSR}}^{s}[\bar{P},\, P,\delta\vec D^s]) \\
   & & \times\exp\left(\int d\tau d^{N}n J(\tau,\vec{n})P(\tau,\vec{n})\right)\exp\left(-\int d^{N}n\frac{(\delta \vec D^{s})^{2}}{\Delta}\right) \ , \nonumber
\eea
where
\be
S_{{\rm MSR}}^{s}[\bar{P},\, P,\,\delta\vec D^s]=i\int d\tau d^{N}n\,\bar{P}\left(\frac{\partial P}{\partial\tau}-\sum_d^N\frac{\partial}{\partial n^d}\left((D+\delta D_d^s)\frac{\partial P}{\partial n^d}\right)\right) \ .
\ee
Since the disorder is Gaussian, we can readily integrate it
out to find
\be
Z[J]=\int{\cal D}\bar{P}{\cal D}P\exp(S_{{\rm eff}}^{s}[\bar{P},\, P])\exp\left(\int d\tau d^{N}n\, J(\tau,\vec{n})P(\tau,\vec{n})\right) \ ,
\ee
where the effective action is
\bea
S_{eff}^{s} & = & i\int d\tau d^{N}n\,\bar{P}\left(\frac{\partial P}{\partial\tau}-D\triangle P\right) \label{eq:EffActionRandomDiffusion}\\
 & & -\frac{\Delta}{2}\sum_{d,e}^N\int d\tau d\tau'd^{N}n\frac{\partial\bar{P}(\tau,\vec{n})}{\partial n_d}\frac{\partial P(\tau,\vec{n})}{\partial n_d}\frac{\partial\bar{P}(\tau',\vec{n})}{\partial n_e}\frac{\partial P(\tau',\vec{n})}{\partial n_e} \ . \nonumber
\eea
Simple scaling analysis (analogous to that of Section \ref{sec:Random-environment}) shows that $\Delta=0$ is a fixed point of the renormalization group flow. Rescaling the
length $\vec{n}$ by a factor of $l$ and requiring that the bare propagator
stays invariant, will allow us to verify that the dynamical exponent
(the one which determines how the frequency $\omega'=\omega l^{z}$
is rescaled) is again $z=2$, and the rescaling law for Fourier components
of the MSR fields is the same as we found in Section \ref{sec:Random-environment}. Disorder strength is
renormalized as $\Delta'=\Delta l^{-N}$. We therefore conclude that a weak disorder in anisotropic diffusion
coefficients is irrelevant for any $N$. One can confirm this conclusion by applying exact renormalization
group analysis.

%%%%%%%%%%%%%%%%%%%%%%%%%%%%%%%%%%%%%%%%%%%%%%%%%%%%%%%%%%%%%%%%%%%%%%%%%%%%%%%%%%%%%%%%
%%%%%%%%%%%%%%%%%%%%%%%%%%%%%%%%%%%%%%%%%%%%%%%%%%%%%%%%%%%%%%%%%%%%%%%%%%%%%%%%%%%%%%%%
%%%%%%%%%%%%%%%%%%%%%%%%%%%%%%%%%%%%%%%%%%%%%%%%%%%%%%%%%%%%%%%%%%%%%%%%%%%%%%%%%%%%%%%%
%%%%%%%%%%%%%%%%%%%%%%%%%%%%%%%%%%%%%%%%%%%%%%%%%%%%%%%%%%%%%%%%%%%%%%%%%%%%%%%%%%%%%%%%
%%%%%%%%%%%%%%%%%%%%%%%%%%%%%%%%%%%%%%%%%%%%%%%%%%%%%%%%%%%%%%%%%%%%%%%%%%%%%%%%%%%%%%%%

\subsection{Random isotropic diffusion: Hermitian case}\label{sec:HermitianCase}

According to the renormalization group analysis, random fluctuations
of the anisotropic diffusion coefficients $\delta D_d^s$
do not affect the character of diffusion of the probability distribution
$P(\vec{n},\tau)$ in the limit $\tau\to\infty$ for arbitrary $N$. This, however, does not mean that the dynamics of the probability distribution
on the landscape is not influenced by disorder $\delta D^s_d$ at intermediate time scales. To
get some idea for what we might expect at intermediate $\tau$, let
us consider the Fokker-Planck equation
\be
 \frac{\partial P}{\partial\tau} = \hat{H}P \label{eq:FPAnd}
\ee
with the following Hermitian\footnote{With respect to the scalar product $(\Psi ', \Psi ) = \int d^N n \Psi'(\vec{n}) \Psi (\vec{n})$.} Hamiltonian \cite{PodolskyEnqvist}
\be
 \hat{H} = D\triangle+\partial_d(V(\vec{n})\cdot\partial_d) \ , \label{eq:HermiteanHamiltonian}
\ee
where $V(\vec{n})$ is a random Gaussian-distributed potential. This is a special case of (\ref{eq:aniso}) with only the isotropic piece (\emph{i.e.}, independent of $d$-directions) of the anisotropic diffusion coefficients $\delta D_d^s$ non-vanishing.

%\lan{Good in this sentence?} This
%case can be trivially deduced from the one discussed in the
%Section \ref{sec:Anisotropic-diffusion} by considering only an \emph{isotropic} (i.e., independent of
%$d$-directions) part of the anisotropic diffusion coefficients $\delta D_d^s$.

%The crucial difference between isotropic and general cases is
%that
For Hermitian $\hat{H}$ the solution of the Fokker-Planck
equation (\ref{eq:FPAnd}) can be expanded  in terms of a complete set of orthogonal
eigenfunctions of the Hamiltonian (\ref{eq:HermiteanHamiltonian}), and all corresponding eigenvalues will be real.\footnote{Let us return from the continuum limit \refb{eq:DiffusionEquation1D} back to the discrete set of the vacuum dynamics equations \refb{VDlatticized}, where the matrix $H_{ij}$ has a general form, \emph{i.e.,} it is neither normal nor (what is even more restrictive) Hermitian. In this case, the set of eigenfunctions of the operator $H_{ij}$ may be incomplete.}
The solution of (\ref{eq:FPAnd}) has the form (below we take $V\ll D$, but the generalization
for arbitrary $V$ is straightforward)
\be
 P(\vec{n},\tau) = \sum_{k=0}^\infty c_{k}\psi_{k}(\vec{n})e^{-E_{k}(\tau-\tau_{0})} \ , \label{eq:FokkerPlanckSolution}
\ee
where $\psi_{k}$ and $E_{k}$ are eigenfunctions and eigenvalues
of the following Schr\"odinger equation
\be
 \partial_{n}^{2}\psi_{k}+\left(\frac{E_{k}}{\sqrt{D}}-W(\vec{n})\right) = 0 \label{eq:SchrodingerHermitian}
\ee
and the effective potential $W(\vec n)$ is given by
\be
 W(\vec{n})=\frac{(\partial_{n}V)^{2}}{4D^{2}}-\frac{\partial_{n}^{2}V}{2D} \ . \label{eq:Superpotential}
\ee
Since the effective potential $W(\vec{n})$ in (\ref{eq:SchrodingerHermitian})
has a supersymmetric form, all eigenvalues $E_{n}$ are positive definite.
If the ``volume'' $\int d^{N}n$ of the moduli space of the landscape  is finite, the ground state $\psi_0$ (normalized to $\psi_{0}(\vec{n})=1$) is trivial and has zero energy $E_0 = 0$.  As $\tau\to\infty$
(or more accurately, at $\tau\gg E_{1}^{-1}$) only the first term
corresponding to the trivial ground state survives in the expansion
(\ref{eq:FokkerPlanckSolution}).

Let us now average over disorder, apply the RG machinery used
in previous Sections and reproduce this asymptotic behavior. After introducing Martin-Siggia-Rose functional and integrating out the Gaussian disorder
\be
 \langle V(\vec{n})V(\vec{n}')\rangle = \mu\delta(\vec n-\vec n') \ ,
\ee
where $\mu$ is a constant, of the random potential $V(\vec{n})$ we find the effective action
of the form similar to (\ref{eq:EffActionRandomDiffusion}). Scaling
analysis again shows that disorder of $V(\vec{n})$ is irrelevant
for arbitrary effective number of dimensions $N$. The physical reason for this
is now clear: the Hamiltonian (\ref{eq:HermiteanHamiltonian}) has
a trivial extended eigenstate with $E_{0}=0$ defining asymptotic
behavior of the probability distribution as $\tau\to\infty$.

Late time (large but finite $\tau$) behavior is determined by eigenstates $\psi_{k}(\vec{n})$
with higher $k>0$. Also, if the ground state $\psi_{0}(\vec{n})$
is not normalizable (such as in the case of an infinite landscape
of vacua), higher $k$ eigenstates are relevant in determining the %$\tau\to\infty$
late time asymptotics.

The following picture \cite{PhononsLocalization} emerges
from the analysis of the nonlinear sigma model corresponding to the
Martin-Siggia-Rose functional (\ref{eq:EffActionRandomDiffusion}).
For $N=1$ (quasi-one-dimensional islands on the landscape) all eigenstates
$\psi_{k}(n)$ with $k>0$ are localized near the so called localization
centers $n=n_{i}$ in the sense that wave functions $\psi_{k}$ are
exponentially peaked at $n=n_{i}$. Localization centers correspond
to the lowest local minima of the superpotential $W(\vec{n})$.

The width of $\psi_{k}$ is called the localization length $\xi_{k}$.
At $N=1$ it behaves as
\be
 \xi_{k}\sim E_{k}^{-1}\sim k^{-2} \ ,
\ee
\emph{i.e.}, it grows at small energies and reaches $\infty$ for $E=E_{0}=0$.
For small $\tau$, the localized eigenstates with higher energy have significant contribution to the probability distribution (\ref{eq:FokkerPlanckSolution}), thus affecting the character of diffusion.
%Contribution of higher energy localized eigenstates into the expression
%for the probability distribution (\ref{eq:FokkerPlanckSolution}) is
%large for small $\tau$ so that localized states influence the character
%of diffusion.

\begin{figure}
\begin{center}
\includegraphics[scale=0.8]{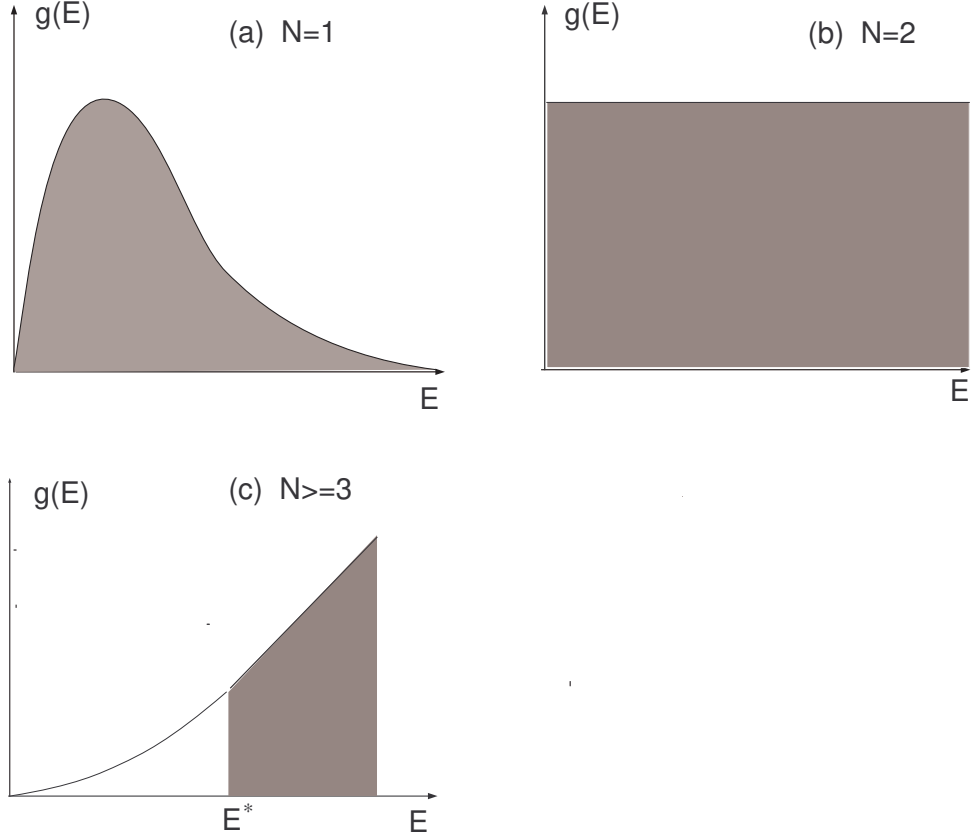}
\caption{Densities of states $g(E)$ for $N=1$, $N=2$ and $N\ge3$. In the
$N=1$ case (a) all states $\psi(E,n)$ are localized (gray region); discreteness
of the spectrum should be taken into account for small $E$. Also,
in the $N=2$ case (b) all states are localized, and the density of states
remains constant. Finally, for $N\ge3$ (the case (c)) there exists a mobility edge
$E^{*}$ such that all states with $E>E^{*}$ are localized, while
states with $E<E^{*}$ are not.\label{fig:density}}
\end{center}
\end{figure}

Behavior of the correlation function at intermediate times can be
estimated as follows. For simplicity, let us suppose that all eigenstates
are localized near the same localization center $n_{i}$ (\emph{e.g.}, in
the case when a single very deep minimum of the superpotential
$W(n)$ is relevant), so that
\be
 \psi_k(n) \simeq \exp\left(-\frac{|\vec{n}-\vec{n}_{i}|}{\xi_{k}}\right) \ .
\ee
We therefore have
\bea
\langle(\vec{n}-\vec{n}_{i})^{2}\rangle & \simeq & \frac{\int d^{N}l\, d^{N}n\,(\vec{n}-\vec{n}_{i})^{2}e^{-\frac{|\vec{n}-\vec{n}_{i}|}{\xi(l)}-\sum_{k}E_{1}l_{k}^{2}\tau}}{\int d^{N}l\, d^{N}n\, e^{-\frac{|\vec{n}-\vec{n}_{i}|}{\xi(l)}-\sum_{k}E_{1}l_{k}^{2}\tau}} \\
 & = & \frac{\int dE\, g(E)\,\xi^{N+2}(E)e^{-E\tau}}{\int dE\, g(E)\,\xi^{N}(E)e^{-E\tau}} \ ,
\eea
where $E_1$ is the energy of the first excited state\footnote{It is inversely proportional to the ``volume'' of the moduli space of the island and determines the longest time scale in the problem.} and $g(E)$ is the density of states and it is proportional to
\be
 g(E) \propto E^{\frac{N-2}{2}} \ .
\ee
For $N=1$ and $\tau\ll E_{1}^{-1}$ we find
\be
 \langle(n-n_{i})^{2}\rangle\sim\tau^{2} \ .
\ee
This means that the diffusion is slowed down in one dimension due to the presence
of localized states. The slow-down is not very strong, since the density
of states grows with decreasing energy. The density of states $g(E)$ is depicted in Fig. \ref{fig:density} for arbitrary $N$.

Similarly for $N=2$ (quasi-two-dimensional case), all eigenstates
$\psi_{k}(\vec{n})$ with $k>0$ are localized near localization
centers but the localization length now grows much faster as $E\to0$
\be
 \xi(E)\sim e^{1/E} \ .
\ee
However, the density of states now remains constant as $E \to 0$,
so diffusion slows down more effectively.

Finally, for $N>2$ there exists a mobility edge $E^{*}$ such
that all states with $E_{k}<E^{*}$ are extended, while states with
$E_{k}>E^{*}$ are localized. The localization length in this case behaves as
\be
 \xi(E) \sim \frac{1}{(E-E^{*})^{\frac{1}{N-2}}}
\ee
near the threshold $E=E^{*}$. Linear diffusion asymptotics
\be
 \langle\vec{n}^{2}(\tau)\rangle \sim \tau
\ee
is also reproduced when $\tau\gg(E^{*})^{-1}\gg E_{1}^{-1}$. The diffusion process of the probability distribution is less effective
at $(E^{*})^{-1}\gg \tau\gg E_1^{-1}$: mean-square displacement behaves as
\be
 \langle(\vec{n}(\tau)-\vec{n}_i)^{2}\rangle \sim \tau^{\frac{2}{N-2}} \ .
\ee

%%%%%%%%%%%%%%%%%%%%%%%%%%%%%%%%%%%%%%%%%%%%%%%%%%%%%%%%%%%%%%%%%%%%%%%%%%%%%%%%%%%%%%%%
%%%%%%%%%%%%%%%%%%%%%%%%%%%%%%%%%%%%%%%%%%%%%%%%%%%%%%%%%%%%%%%%%%%%%%%%%%%%%%%%%%%%%%%%
%%%%%%%%%%%%%%%%%%%%%%%%%%%%%%%%%%%%%%%%%%%%%%%%%%%%%%%%%%%%%%%%%%%%%%%%%%%%%%%%%%%%%%%%
%%%%%%%%%%%%%%%%%%%%%%%%%%%%%%%%%%%%%%%%%%%%%%%%%%%%%%%%%%%%%%%%%%%%%%%%%%%%%%%%%%%%%%%%
%%%%%%%%%%%%%%%%%%%%%%%%%%%%%%%%%%%%%%%%%%%%%%%%%%%%%%%%%%%%%%%%%%%%%%%%%%%%%%%%%%%%%%%%

\section{Conclusions and discussion}\label{sec:conclusions}

In this paper, we systematically study dynamics of eternal inflation on the string theory landscape in the presence of disorder.
Due to a non-trivial dynamics of moduli fields, many sources of disorder may exist in different sectors of the landscape. We identify one such sector with built-in disorder in Section \ref{sec:eternalKS}. This sector is established by compactifying Type IIB string theory on a class of Calabi-Yau manifolds with a number of warped Klebanov-Strassler throats attached to the bulk part of the Calabi-Yau (see Fig. \ref{fig2}). KS throats can be attached \emph{randomly} to the bulk CY and can have \emph{random} lengths.

Eternal inflation on such sector of the landscape has a mechanism different from the one realized on the Bousso-Polchinski sector \cite{PolchinskiBousso}.
Our four-dimensional Universe is localized on a D3-brane located in
one of the KS throats, and effective value of the cosmological constant
for the four-dimensional observer living on the brane is essentially determined by the distance
$r$ between the D3-brane and the tip of the KS
throat (more accurately, this distance determines the value of the
effective inflaton field $\phi=\sqrt{T_{3}}r$ detected by the four-dimensional
observer). The D3-brane travels %towards the tip of
in the throat, as a result, four-dimensional
Hubble patch, the one our observer lives in, inflates -- this is the KKLMMT scenario \cite{KKLMMT} in a nutshell.
Eternal inflation (as we proved in Section \ref{sec:eternalKS}, it is possible
in a KKLMMT setup) corresponds to strong fluctuations of the
inflaton field, while the distance $r$ of the D3-brane from the
tip of the throat is large. The D3-brane can leave the given KS
throat and enter other KS throats, with different lengths, \emph{i.e.}, different
scales of the cosmological constant measured by a four-dimensional observer.
%living in a given four-dimensional Hubble patch.
By construction, KS throats (with random lengths) are randomly attached to the bulk CY, and this disorder influences dynamics of eternal inflation.

After identifying a possible source of disorder on the string theory landscape, we consider the vacuum dynamics equations (\ref{eq:VacuumDynamicsNonVolume}) describing dynamics of eternal inflation on a disordered sector of the landscape (or entire landscape). One particular goal of our study is to understand how disorder influences the %late time ($\tau \to \infty$) asymptotics and
late time ($\tau$ large but finite) behavior of the probability $P_i (\tau )$ to measure a given value of the cosmological constant $\Lambda_i$ in a given four-dimensional Hubble patch.

Discussing eternal inflation on the landscape, usually one is interested in behavior of $P_i (\tau)$ at $\tau \to \infty$. If the set of eigenfunctions of the operator $\hat{H}$ is complete, $P_i (\tau \to \infty)$ behavior is determined by the eigenvalue $E_0$ of $\hat{H}$ with lowest possible real part. One notes that it always exists since the spectrum of $\hat{H}$ is bounded from below for the compact landscape. Generally, $E_0$ is complex, and its real part can be negative. If this is so, $P(\vec{n},\tau )$ itself as well as correlation functions weighted over this probability distribution are never gauge-invariant, since they explicitly depend on $\tau$. If the operator $\hat{H}$ is Hermitian and supersymmetric (such as for the case discussed in the Section \ref{sec:HermitianCase}), the ground state corresponds to $E_0 = 0$, and the asymptotics $P(\tau \to \infty)$ is stationary. For the case of isotropic diffusion discussed in the Section \ref{sec:HermitianCase} the form of the lowest eigenstate is trivial (constant). At weak disorder behavior of $P(\tau = \infty)$ does not depend on initial conditions for eternal inflation on such a landscape, on reparametrisations of $\tau$ and disorder (see the Section \ref{sec:Random-environment}).

%In the present publication, our goal was to determine behavior of $P_i (\tau )$ at large but finite $\tau$.
Although it is physically important to know $P_i (\tau = \infty)$ asymptotics on the landscape,
finite time behavior of $P_i (\tau)$ is also of great physical relevance. Indeed,
because the moduli space of the string theory landscape is huge albeit compact, the late time asymptotics $P_i (\tau = \infty )$ is reached at time scales roughly estimated as $\tau \gg \tau_1 \sim n_{\rm max} $, where $n_{\rm max}$ is the number of vacua on the landscape, \emph{i.e.}, the volume of its compact moduli space. This time scale is extremely long, so it is plausible that our own Hubble patch was generated in the epoch $\tau \ll \tau_1 \sim n_{\rm max}$.\footnote{In this respect, let us again emphasize that we study behavior of the non-volume-weighted measure at $\tau\to\infty$. Behavior of the volume-weighted measure is very different: for example, it is much more likely that our Hubble patch is generated at late times if probabilities are given by integrating over volume-weighted measure.}

%Moreover, late time asymptotics $P_i (\tau = \infty )$ independent on initial conditions for eternal inflation %\emph{may not exist}. Two known examples include solutions of vacuum dynamics equations with volumes of inflating %Hubble patches and with AdS sinks taken into account. In this paper we presented another example; a landscape with %large disordered antisymmetric part of the tunneling rates
%\be
%\delta \Gamma^a_{ij} = \frac{1}{2}\left( \Gamma_{ij} - \Gamma_{ji} \right) \ .
%\ee
%In these three cases, the probability flow is not unitary in the sense that the operator $\hat{H}$ in the continuum %version of the vacuum dynamics equations
%\be
%\frac{\partial P (\vec{n} , \tau)}{\partial \tau} = \hat{H} P (\vec{n}, \tau)
%\label{VDEconcl}
%\ee
%is not Hermitian, and behavior of $P_i (\tau)$ at $\tau \to \infty$ is \emph{always} determined by its intermediate %time asymptotics. Using an explicit example of the landscape with non-Hermitian $\hat{H}$ we demonstrated
%that it is actually clear that typically, \emph{i.e.}, on a generic sector of the landscape with disorder, late %time asymptotics $P_i (\tau = \infty )$ does not exist, and one has to know intermediate time behavior of the %probability distribution $P_i$.

To determine behavior of $P_i (\tau)$ at large but finite $\tau$ in the presence of disorder on the landscape, our strategy was to first construct continuum limit of the vacuum dynamics equations (\ref{eq:VacuumDynamicsNonVolume}) and then determine the asymptotic behavior of the probability distribution $P_i (\tau )$ %followed
by applying dynamical renormalization group methods.

It is easy to understand why studying the continuum limit of the equations (\ref{eq:VacuumDynamicsNonVolume}) is appropriate if we consider a landscape with Hermitian $\hat{H}$. In this case, dynamics of $P_i(\tau)$ is completely determined by real eigenvalues and the form of eigenstates of the operator $\hat{H}$ (see the Subsection \ref{sec:HermitianCase}). At intermediate time scales $\tau \ll E_1^{-1}$ (here $E_1$ is the first excited eigenstate) the spectrum of $\hat{H}$ can be approximately considered continuous since the energy gap between eigenstates is of the order of inverse volume $1/n_{\rm max}$ of the moduli space of the landscape (or the part of the landscape under consideration).

The use of dynamical RG methods is appropriate once recalling why RG is applicable in a generic quantum field theory (QFT): in $k\to 0$ limit only renormalizable interactions or, in other words, only relevant terms in the effective action of the QFT survive. Similarly, for a time-dependent problem only relevant terms survive in the effective Martin-Siggia-Rose action \cite{MartinSiggiaRose} describing the dynamics of $P_i (\tau)$ (see Section \ref{sec:Disorder}).

The results of our weak disorder analysis are the following. In presence of disorder on the landscape

\begin{enumerate}
\item large $\tau$ behavior of the probability distribution $P_i(\tau)$ is strongly modified on islands of the landscape with $N<3$, where $N$ is the Hausdorff dimension of the tunneling graph corresponding to the given island. There exists a non-trivial fixed point in the RG flow corresponding to finite disorder (see Subsection \ref{sec:Random-environment}) which defines this anomalous behavior.
\item large $\tau$ behavior of the probability distribution $P_i(\tau)$ for islands with $N>3$ is not affected by weak disorder since the only fixed point of the RG flow corresponds to the absence of any disorder (again see the Subsection \ref{sec:Random-environment}).
\item on sectors of the landscape, where $\hat{H}$ is Hermitian, finite $\tau$ behavior is \emph{always} affected by weak disorder for arbitrary $N$, and diffusion of the probability $P_i(\tau)$ is generally suppressed for any $N$. The reason of this suppression is the existence of localized eigenstates of the operator $\hat{H}$, \emph{i.e.,} Anderson localization on sectors of the landscape with disorder.
%\item anomalous $\tau \to \infty$ behavior on the islands with arbitrary (non-Hermitian) $\hat{H}$ cannot be %explained by Anderson localization of the $\hat{H}$ eigenstates, since the general solution of (\ref{VDEconcl}) %\emph{cannot} be represented as a sum over complete set of orthogonal eigenstates of the operator $\hat{H}$.
\end{enumerate}

The situation in the presence of strong disorder on the string theory landscape is much harder to understand qualitatively. Infinitely strong disorder corresponds to a fixed point of the renormalization group flow
of the MSR action. The physics of infinite disorder case
is such that the diffusion of the probability distribution function $P_i(\tau)$ is suppressed completely: as $\tau\to\infty$, the mean-square displacement $\langle n^2 (\tau) \rangle$ approaches a constant for arbitrary $N$.
This fixed point is believed to be unstable with respect to inverse disorder corrections. At $N=1$ this follows from the Sinai's result \cite{Sinai} showing that the mean-square displacement diverges as $\tau\to\infty$
for large but finite disorder. Any proof of this statement is absent for islands with $N>1$.

%%%%%%%%%%%%%%%%%%%%%%%%%%%%%%%%%%%%%%%%%%%%%%%%%%%%%%%%%%%%%%%%%%%%%%%%%%%%%%%%%%%%%%%%
%%%%%%%%%%%%%%%%%%%%%%%%%%%%%%%%%%%%%%%%%%%%%%%%%%%%%%%%%%%%%%%%%%%%%%%%%%%%%%%%%%%%%%%%
%%%%%%%%%%%%%%%%%%%%%%%%%%%%%%%%%%%%%%%%%%%%%%%%%%%%%%%%%%%%%%%%%%%%%%%%%%%%%%%%%%%%%%%%
%%%%%%%%%%%%%%%%%%%%%%%%%%%%%%%%%%%%%%%%%%%%%%%%%%%%%%%%%%%%%%%%%%%%%%%%%%%%%%%%%%%%%%%%
%%%%%%%%%%%%%%%%%%%%%%%%%%%%%%%%%%%%%%%%%%%%%%%%%%%%%%%%%%%%%%%%%%%%%%%%%%%%%%%%%%%%%%%%

\bigskip
\bigskip

\noindent
{\bf \large Acknowledgments}

\bigskip

We would like to thank L. Kofman, S.-H. Henry Tye, A. Vilenkin and especially S. Winitzki for useful comments.
J.M. would like to thank all the members and in particular Esko Keski-Vakkuri of String Theory and Astrophysics group of the Helsinki Institute of Physics (HIP), University of Helsinki for their generosity, hospitality and support, where this project was started when he was a postdoctoral fellow and the people of India for generously supporting research in string theory. N.J. has been in part supported by the Magnus Ehrnrooth foundation. This work was also partially supported by the EU 6th Framework Marie Curie Research and Training network ``UniverseNet'' (MRTN-CT-2006-035863).

\bigskip

\bigskip

%%%%%%%%%%%%%%%%%%%%%%%%%%%%%%%%%%%%%%%%%%%%%%%%%%%%%%%%%%%%%%%%%%%%%%%%%%%%%%%%%%%%%%%%
%%%%%%%%%%%%%%%%%%%%%%%%%%%%%%%%%%%%%%%%%%%%%%%%%%%%%%%%%%%%%%%%%%%%%%%%%%%%%%%%%%%%%%%%
%%%%%%%%%%%%%%%%%%%%%%%%%%%%%%%%%%%%%%%%%%%%%%%%%%%%%%%%%%%%%%%%%%%%%%%%%%%%%%%%%%%%%%%%
%%%%%%%%%%%%%%%%%%%%%%%%%%%%%%%%%%%%%%%%%%%%%%%%%%%%%%%%%%%%%%%%%%%%%%%%%%%%%%%%%%%%%%%%
%%%%%%%%%%%%%%%%%%%%%%%%%%%%%%%%%%%%%%%%%%%%%%%%%%%%%%%%%%%%%%%%%%%%%%%%%%%%%%%%%%%%%%%%

\appendix

\section{Why average over disorder?}\label{sec:whyaverage}

In this Appendix we will explain why it is necessary to average over disorder to
determine the late time dynamics of the probability distribution, giving us a way
to measure a given value of the cosmological constant in a given Hubble
patch. %Nature should choose a single particular realization
%of this disorder.
To understand the physical meaning of averaging over
disorder, let us consider a particular realization of the disorder
on the quasi-one-dimensional ($N=1$) landscape. This system is described by the continuous
version of the vacuum dynamics equations
\be
 \frac{\partial P}{\partial\tau} = \hat H P \ , \label{eq:FPAppA}
\ee
where the operator $\hat{H}$ is Hermitian.\footnote{For $N>1$ this is only possible
if the coefficients $\delta D_d^a=0$, while the diffusion
is isotropic, \emph{i.e.}, $\delta D_d^s=\delta D\cdot\hat{I}$, where $I={\rm diag} (1,1,\ldots)$;
the latter condition is always realized on the quasi-one-dimensional
island.} In this case, the solution of (\ref{eq:FPAppA})
can be represented as a sum over complete set of orthogonal eigenstates
of the operator $\hat{H}$.

Since the moduli space of the island is compact, the distribution
function $P(n,\tau)$ reaches its equilibrium asymptotics
$P_{0}(n)$ in finite time $\tau\sim E_{1}^{-1}$, where $E_{1}$
is the eigenvalue of the first excited eigenstate of
the operator $\hat{H}$. The asymptotics $P_0(n)$ can be solved from the equation
\be
 \frac{\partial}{\partial n}\left((D+\delta D^{s}(n))\frac{\partial P_{0}(n)}{\partial n}\right) = 0 \ . \label{eq:TimeIndependentAsymptoticsEquation}
\ee
The solution does not depend on $n$ nor disorder and it reads
\be
 P_{0}(n)=\frac{1}{\int dn} = \frac{1}{n_{{\rm max}}} \ .
\ee

The general solution of (\ref{eq:FPAppA}) has the form
\be
 P(n,\tau) = \sum_{k}c_{k}\phi_{k}(n)e^{-E_{k}\tau} \ ,
\ee
where $\phi_{k}$ and $E_{k}$ are given by the Schr\"odinger
equation
\be
 D\phi_{k}''+(\delta D^{s}(n)\phi_{k}')'=E_{k}\phi_{k} \ .
\ee
The coefficients $c_{k}$ are defined by the initial conditions for
the probability distribution $P(n,\tau=0)$.

If the disorder on the landscape is weak, \emph{i.e.}, $\delta D^{s}\ll D$,
the behavior of the probability distribution $P(n,\tau)$ at finite
time can be determined by means of perturbation theory. Since the
moduli space of the given island is compact, the spectrum of eigenvalues
$E_{k}$ is discrete. In the leading approximation (\emph{i.e.}, in absence
of any disorder), the spectrum of eigenstates is given by
\be
 E_{k}^{(0)} = \frac{D\pi^{2}k^{2}}{n_{{\rm max}}^{2}} \ ,
\ee
where $n_{\rm max}$ is the maximum allowed index on the island (\emph{i.e.}, volume
of its moduli space). The eigenfunctions are
\be
|k\rangle=\phi_{k}^{(0)}(n)=c_{k}\cos\left(\frac{kn}{n_{{\rm max}}}\right) \ .
\ee
In the first subleading approximation, disorder gives rise to a spectrum shift
\be
 \delta E_{k}^{(1)} = \langle k|\delta\hat{H}|k\rangle\equiv\int dn\, c_{k}^{2}\cos\left(\frac{kn}{n_{{\rm max}}}\right)\delta\hat{H}\cos\left(\frac{kn}{n_{{\rm max}}}\right)  \label{eq:SchrodingerECorrection}
\ee
and the following modification in the form of eigenstates (we suppose that the spectrum of states is not degenerate; the generalization for the case with degeneracy is trivial)
\be
 \delta\phi_{k}^{(1)} = \sum_{k'\ne k}\frac{\langle k'|\delta\hat{H}|k\rangle}{E_{k'}^{(0)}-E_{k}^{(0)}}\phi_{k'}^{(0)} \ , \label{eq:SchrodingerPhiCorrection}
\ee
where the interaction Hamiltonian is
\be
\delta\hat{H}\phi_{k} = \frac{\partial}{\partial n}\left(\delta D^{s}(n)\frac{\partial}{\partial n}\phi_{k}\right) \ .
\ee
We find
\bea
 \delta E_{k}^{(1)}  & = & -c_{k}^{2}\frac{k^{2}}{n_{{\rm max}}^{2}}\int dn\,\delta D^{s}(n)\sin^{2}\left(\frac{kn}{n_{{\rm max}}}\right) \label{eq:Ecorr} \\
\delta\phi_{k}^{(1)} & = & \sum_{k'\ne k}\frac{c_{k}c_{k'}^2\int dn'\, k\, k'\delta D^{s}(n)\sin\left(\frac{kn'}{n_{{\rm max}}}\right)\sin\left(\frac{k'n'}{n_{{\rm max}}}\right)}{D\pi^2(k^{2}-k'^{2})}\cos\left(\frac{k'n}{n_{{\rm max}}}\right) \ . \label{eq:phicorr}
\eea

The main conclusion we draw is the following. As we can see from
the expressions (\ref{eq:Ecorr}) and (\ref{eq:phicorr}), after taking
small disorder on the landscape into account, the probability distribution
$P(n,\tau)$ acquires small corrections having the form of integrals
of disorder over the moduli space of the island, \emph{i.e.}, averaging
of one particular realization of disorder over the overall volume
of the moduli space. By ergodicity, this is equivalent to averaging
over all possible realizations of the disorder.

%%%%%%%%%%%%%%%%%%%%%%%%%%%%%%%%%%%%%%%%%%%%%%%%%%%%%%%%%%%%%%%%%%%%%%%%%%%%%%%%%%%%%%%%
%%%%%%%%%%%%%%%%%%%%%%%%%%%%%%%%%%%%%%%%%%%%%%%%%%%%%%%%%%%%%%%%%%%%%%%%%%%%%%%%%%%%%%%%
%%%%%%%%%%%%%%%%%%%%%%%%%%%%%%%%%%%%%%%%%%%%%%%%%%%%%%%%%%%%%%%%%%%%%%%%%%%%%%%%%%%%%%%%
%%%%%%%%%%%%%%%%%%%%%%%%%%%%%%%%%%%%%%%%%%%%%%%%%%%%%%%%%%%%%%%%%%%%%%%%%%%%%%%%%%%%%%%%
%%%%%%%%%%%%%%%%%%%%%%%%%%%%%%%%%%%%%%%%%%%%%%%%%%%%%%%%%%%%%%%%%%%%%%%%%%%%%%%%%%%%%%%%

\section{Schwinger-Keldysh diagrammatic technique}\label{sec:keldysh}

In the Appendices \ref{sec:keldysh} and \ref{sec:MSR} we mostly follow excellent lecture notes
by A. Kamenev \cite{Kamenev} and the classical book by L. Kadanoff
and G. Baym \cite{KadanoffBaimBook}, where the reader can find
a much deeper overview of the Schwinger-Keldysh diagrammatic methods.

\begin{figure}[t]
\includegraphics[scale=0.75]{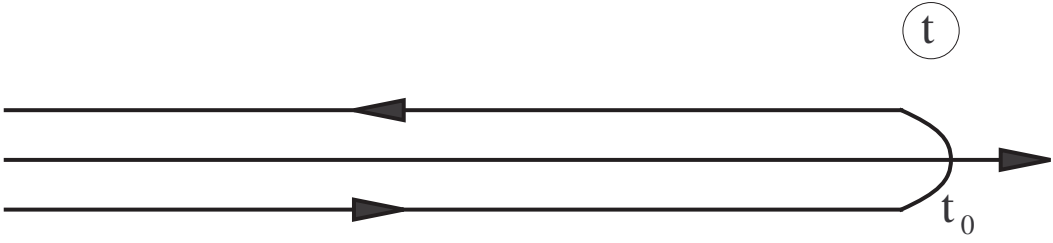}
\caption{The closed time contour for calculating Green functions in the Schwinger-Keldysh
formalism.\label{fig:contour}}

\end{figure}

The Schwinger-Keldysh diagrammatics is used to deal with quantum fields
out of equilibrium. The crucial difference between the equilibrium diagrammatic
(Feynman) methods and the Schwinger-Keldysh diagrammatics is that amplitudes
and partition functions in the latter case are calculated along a
closed time contour in the complex plane. The contour starts from $t=-\infty$, goes
to some fixed moment of time $t=t_{0}$ and then going back to
the infinite past (see Fig. \ref{fig:contour}).\footnote{It is not necessary for the contour to start from infinity. It is actually not suitable to use such a contour for many physical situations
\cite{BergesIntro}. However, using an infinitely long contour is
more convenient, since one keeps the preparation of the
initial state well under control by adiabatically slowly turning on the
interaction term.}

For simplicity, let us suppose that we have a theory with a single real
scalar field $\chi(x)$. The major consequence of calculating amplitudes
along the closed time contour is an effective doubling of degrees of
freedom. In Schwinger-Keldysh diagrammatic technique, instead of a
single field $\chi(x)$ we introduce two \emph{Keldysh fields} $\chi^{+}(x)$
and $\chi^{-}(x)$ defined on the lower and upper sides of the closed
time contour. Because the contour is connected, the fields $\chi^+,\chi^-$ are not independent.
In particular, the correlator $\langle\chi^{+}(x)\chi^{-}(x)\rangle$ is non-vanishing.

After introducing the Keldysh indices $a,b=\{+,-\}$, the generating
functional for the Green functions has the form
\be
 Z[J_{\chi}^{c}]=\int{\cal D}\chi^{a}\exp\left[i\left(S[\chi^{a}]+\int d^{4}x\sqrt{-g(x)}J_{\chi}^{a}(x)\chi^{a}(x)\right)\right] \ . \label{eq:GenFunctional}
\ee
The correlation functions of the fields $\chi^a$ are given by functional derivatives with respect to sources at $J^a_\chi=0$. For example, the two-point
correlation function of $+$ and $-$ Keldysh fields reads
\be
 iG^{+-}(x,x') \equiv \langle\chi^{+}(x)\chi^{-}(x')\rangle = \frac{1}{2}\frac{\delta^{2}Z[J^{+}(x),J^{-}(x')]}{\delta J^{+}(x)\delta J^{-}(x')}\Big|_{J^{+},J^{-}=0} \ .
\ee

Not all of the four possible two-point Green functions\footnote{\emph{I.e.}, the Green functions $iG^{++}=\langle\chi^{+}(x)\chi^{+}(x')\rangle$,
$iG^{--}=\langle\chi^{-}(x)\chi^{-}(x')\rangle$, $iG^{-+}=\langle\chi^{-}(x)\chi^{+}(x')\rangle$
and $iG^{+-}=\langle\chi^{+}(x)\chi^{-}(x')\rangle$.} are independent. Due to causality constraints (see \cite{KadanoffBaimBook}) we have an identity
\be
 G^{++}(x,y)+G^{--}(x,y)=G^{+-}(x,y)+G^{-+}(x,y) \ . \label{eq:CasualIdentity}
\ee
It is therefore possible to simplify the analysis of perturbation
theory by doing the Keldysh rotation
\bea
 \chi_{{\rm cl}}(x) & = & \frac{1}{\sqrt{2}}(\chi^{+}(x)+\chi^{-}(x)) \\
 \chi_{{\rm q}}(x)  & = & \frac{1}{\sqrt{2}}(\chi^{+}(x)-\chi^{-}(x))  \label{eq:KeldyshRotation}
\eea
since only the Green functions $\langle\chi_{{\rm cl}}(x)\chi_{{\rm q}}(x')\rangle\equiv iG^{R}(x,x')$,
$\langle\chi_{{\rm q}}(x)\chi_{{\rm cl}}(x')\rangle\equiv iG^{A}(x,x')$
and $\langle\chi_{{\rm cl}}(x)\chi_{{\rm cl}}(x')\rangle=iG^{K}(x,x')$
are non-zero. $G^{R}(x,x')$ and $G^{A}(x,x')$ are retarded
and advanced Green functions, respectively. $G^{K}$ is called
the Keldysh Green function. The Green function $i\langle\chi_{{\rm q}}(x)\chi_{{\rm q}}(x')\rangle$
remains zero non-perturbatively because the identity (\ref{eq:CasualIdentity})
holds in all orders of the perturbative expansion \cite{Kamenev}.

The field $\chi_{{\rm cl}}(x)$ is usually denoted as {}``classical''
while the field $\chi_{{\rm q}}(x)$ as quantum, since among saddle
points of the effective action there is always a solution such that $\chi_{{\rm q}}(x)=0$
and $\chi_{{\rm cl}}(x)$ satisfies the classical equations of motion
\cite{Kamenev}.

Let us now discuss the information carried by the Keldysh Green functions.
For a free massive scalar field $\chi(x)$ we have
\bea
 G^{++}(k) & = & (k^{2}-m^{2}+i\epsilon)^{-1}-2\pi in(k)\delta(k^{2}-m^{2}) \label{eq:ppgf} \\
 G^{--}(k) & = & -(k^{2}-m^{2}+i\epsilon)^{-1}-2\pi in(k)\delta(k^{2}-m^{2}) \\
 G^{+-}(k) & = & -2\pi i(\theta(k^{0})+n(k))\delta(k^{2}-m^{2}) \label{eq:ppgf3} \ ,
\eea
where $n(k)$ is the occupation number for a given mode with momentum $k$ and $\theta (k)$ is a step function.

The simplest way to verify (\ref{eq:ppgf})-(\ref{eq:ppgf3}) is to use the WKB (Wentzel-Kramers-Brillouin) approximation. For example, the limit $x\to x'$ of the $G^{++}$ Green function
(neglecting the vacuum contribution) is given by
\be
 \langle\chi^{+}(x)\chi^{+}(x)\rangle = \int\frac{d^{3}k}{(2\pi)^{3}2\omega_{k}}n(k)=2\pi\int\frac{d^{4}k}{(2\pi)^{4}}n(k)\delta(\omega_{k}^{2}-k^{2}-m^{2}) \ .
\ee
Recalling that $\langle\chi^{+}(x)\chi^{+}(x)\rangle=iG^{++}(x,x)$
we come to the expression (\ref{eq:ppgf}).

Finally, by performing the Keldysh rotation (\ref{eq:KeldyshRotation}) yields
\bea
 G^{K}(k) & = & -2\pi i(1+2n(k))\delta(k^{2}-m^{2}) \\
 G^{R}(k) & = & (k^{2}-m^{2}+i0)^{-1} \\
 G^{A}(k) & = & (k^{2}-m^{2}-i0)^{-1} \ .
\eea
We conclude that the Keldysh Green function $G^{K}$ carries information
about the distribution function $n(k)$, while the retarded and advanced
Green functions $G^R,G^A$ give the spectrum of particles (and are independent
of the distribution function $n(k)$). This separation is only valid for systems
close enough to a thermal equilibrium.\footnote{Far from equilibrium the situation is not so trivial anymore, but
it is possible to show that imaginary parts of the Green functions
still carry information about the occupation numbers in the corresponding
modes \cite{Kamenev}.}

%%%%%%%%%%%%%%%%%%%%%%%%%%%%%%%%%%%%%%%%%%%%%%%%%%%%%%%%%%%%%%%%%%%%%%%%%%%%%%%%%%%%%%%%
%%%%%%%%%%%%%%%%%%%%%%%%%%%%%%%%%%%%%%%%%%%%%%%%%%%%%%%%%%%%%%%%%%%%%%%%%%%%%%%%%%%%%%%%
%%%%%%%%%%%%%%%%%%%%%%%%%%%%%%%%%%%%%%%%%%%%%%%%%%%%%%%%%%%%%%%%%%%%%%%%%%%%%%%%%%%%%%%%
%%%%%%%%%%%%%%%%%%%%%%%%%%%%%%%%%%%%%%%%%%%%%%%%%%%%%%%%%%%%%%%%%%%%%%%%%%%%%%%%%%%%%%%%
%%%%%%%%%%%%%%%%%%%%%%%%%%%%%%%%%%%%%%%%%%%%%%%%%%%%%%%%%%%%%%%%%%%%%%%%%%%%%%%%%%%%%%%%

\section{Quasi-classical Keldysh action, Martin-Siggia-Rose diagrammatics}\label{sec:MSR}

The trivial saddle point of the generating functional (\ref{eq:GenFunctional})
rewritten in terms of quantum and classical Keldysh fields (\ref{eq:KeldyshRotation})
is determined by the equations
\bea
 \frac{\delta S}{\delta{\chi}_{\rm cl}} & = & 0 \quad \to \quad \chi_{\rm q}=0 \\
 \frac{\delta S}{\delta{\chi}_{\rm q}}  & = & 2{\cal O}_{R}[\chi_{\rm cl}]\chi_{\rm cl}=0 \ ,
\eea
where ${\cal O}_R$ is a retarded operator describing the dynamics
of the classical Keldysh field, while the fields $\chi_{\rm cl},\chi_{\rm q}$ were introduced in the previous Appendix. The trivial saddle point is $S=0$ with $Z=1$.\footnote{It is possible that the Keldysh generating functional can have other saddle points (similar to instanton trajectories in the
equilibrium situation) but they have zero contribution to the partition
function $Z$.} In fact, the condition $Z=1$ holds even if one considers fluctuations
near this saddle point \cite{Kamenev}.

To consider the quasi-classical limit, fluctuations of $\chi_{q}$
should be allowed near the classical trajectory. Let us keep terms only
up to second order in $\chi_{q}$ in the Keldysh action.
The semiclassical action will then take the following form
\be
 S_{scl} = 2\int\int dtdt'\left[{\chi}_{\rm q}[{G}^{-1}]^{K}\chi_{\rm q}+({\chi}_{q}{\cal O}^{R}[\chi_{\rm cl}]\chi_{\rm cl}+c.c.)\right] \ ,
\ee
where $c.c.$ denotes the complex conjugation.

One simple way to treat this semiclassical theory is to use the Hubbard--Stratonovich
transformation introducing the auxiliary stochastic field $\xi(t)$
and decoupling the quadratic term in the quasi-classical equation. One
will find that the resulting action is linear with respect to $\chi_{q}$,
\emph{i.e.}, the integration over $\chi_{q}$ leads to the functional $\delta$--function
and to the stochastic Langevin equation
\be
 {\cal O}^{R}[\phi_{\rm cl}]\phi_{\rm cl}(t)=\xi(t) \ , \label{eq:LangevinSK}
\ee
where
\be
 \langle\xi(t){\xi}(t')\rangle=\frac{i}{2}[{G}^{-1}]^{K}(t,t') \ .
\ee

Another way to deal with the semiclassical theory is to integrate out
the $\chi_{\rm q}$ field, since its contribution to the
action is quadratic. The result is the theory of the classical Keldysh
field
\be
 S[\chi_{\rm cl}]=2\int_{-\infty}^\infty dtdt'~{\chi}_{\rm cl}({\cal O}^{A}[\bar{\chi}_{\rm cl}]{G}^{A})[{ G}^{K}]^{-1}({G}^{R}{\cal O}^{R}[\chi_{\rm cl}])\chi_{\rm cl} \ .
\ee
If non-linearities with respect to $\chi_{\rm cl}$ are neglected, the
theory is a free field theory but with a complicated
propagator. Using first quantized version of this theory, one can
show that dynamics of the probability $P(x,t)$ to find a particle
excitation in the point $x$ at time $t$, is governed by the Fokker-Planck
equation. This is not surprising since the variation of the classical limit of the Schwinger-Keldysh equation
gives the Langevin equation (\ref{eq:LangevinSK}).

In fact, the opposite procedure is possible: one can start from a
Langevin/Fokker-Planck equation (or, generally speaking, with \emph{any}
equation describing diffusive dynamics) and construct a classical
limit of the corresponding Schwinger-Keldysh diagrammatic technique.
This procedure was first introduced by Martin, Siggia and Rose \cite{MartinSiggiaRose}.

Consider the Fokker-Planck equation
\be
 \frac{\partial P}{\partial\tau}=\hat{H}P=D\triangle P+\delta\hat{H}P \ .
\ee
The corresponding ``generating functional'' can be written as a functional $\delta$-function
\be
 Z = \int{\cal D}P\delta\left(\frac{\partial P}{\partial t}-\hat{H}P\right) \ ,
\ee
which, after introducing the auxiliary field $\bar{P}$, acquires the form
\be
 Z = \int{\cal D}P{\cal D}\bar{P}\exp\left(i\int dtd\vec{n}\,\bar{P}\left(\frac{\partial P}{\partial t}-\hat{H}P\right)\right) \ . \label{eq:MartinSiggiaRose1}
\ee
There exists a remarkable similarity between (\ref{eq:MartinSiggiaRose1})
and the Schwinger-Keldysh generating functional describing the quasi-classical
approximation of the quantum non-equilibrium dynamics. For example,
in the Martin-Siggia-Rose diagrammatic technique, the field $\bar{P}$
plays the same role as the quantum field $\chi_{\rm q}$ in the Schwinger-Keldysh
framework.

The diagrammatic expansion of the generating
functional (\ref{eq:MartinSiggiaRose1}) is not of much use, since
we are not interested in correlation functions of $P$ or $\bar{P}$:
 $P$ is itself the probability measure for a random walk/diffusion
process governed by the corresponding Fokker-Planck equation. Observables
in the problem under consideration (such as the mean square displacement
$\langle\vec{n}^{2}(\tau)\rangle$) are related to the physics of
this random walk. However, the correlation functions of $\vec{n}$
can be determined from the correlation functions in the momentum representation
generated by $(\ref{eq:MartinSiggiaRose1})$ by differentiating over
momenta. For example, we have
\be
 \langle\vec{n}^{2}(\tau)\rangle = -\frac{d}{dq^d}\frac{d}{dq^d}\int d\omega e^{-i\omega\tau}G(\omega,\vec{q})|_{\vec{q}=0} \ ,
\ee
where $G(\omega,\vec{q})$ is the Fourier component of the Green function
$\langle PP\rangle$. This is due to the correspondence between the
Langevin equation describing the dynamics of the observable $\vec{n}$
and the Fokker-Planck equation describing the dynamics of the probability
distribution $P$ of the diffusion process.

%%%%%%%%%%%%%%%%%%%%%%%%%%%%%%%%%%%%%%%%%%%%%%%%%%%%%%%%%%%%%%%%%%%%%%%%%%%%%%%%%%%%%%%%
%%%%%%%%%%%%%%%%%%%%%%%%%%%%%%%%%%%%%%%%%%%%%%%%%%%%%%%%%%%%%%%%%%%%%%%%%%%%%%%%%%%%%%%%
%%%%%%%%%%%%%%%%%%%%%%%%%%%%%%%%%%%%%%%%%%%%%%%%%%%%%%%%%%%%%%%%%%%%%%%%%%%%%%%%%%%%%%%%
%%%%%%%%%%%%%%%%%%%%%%%%%%%%%%%%%%%%%%%%%%%%%%%%%%%%%%%%%%%%%%%%%%%%%%%%%%%%%%%%%%%%%%%%
%%%%%%%%%%%%%%%%%%%%%%%%%%%%%%%%%%%%%%%%%%%%%%%%%%%%%%%%%%%%%%%%%%%%%%%%%%%%%%%%%%%%%%%%

\end{document}